\documentclass{jfm}

\newcommand{\pdiff}[2]{\frac{\partial #1}{\partial #2}}
\newcommand{\odiff}[2]{\frac{d #1}{d #2}}
\usepackage{amsmath}
\usepackage{amssymb}
\usepackage{color}

\begin{document}

\newtheorem{lemma}{Lemma}
\newtheorem{corollary}{Corollary}

\newcommand{\NY}[1]{{\color{black} #1}}
\newcommand{\SYYYY}[1]{{\color{black} #1}}
\newcommand{\SY}[1]{{\color{black} #1}}
\newcommand{\SYY}[1]{{\color{black} #1}}
\newcommand{\NYY}[1]{{\color{black} #1}}
\newcommand{\SYYY}[1]{{\color{black} #1}}
\newcommand{\SSY}[1]{{\color{black} #1}}
\newcommand{\FSY}[1]{{\color{black} #1}}

\shorttitle{
Collision between chemically driven self-propelled drops
 } 
\shortauthor{S. Yabunaka and N. Yoshinaga} 

\title{Collision between \NY{chemically driven} self-propelled drops}

\author
 {
 Shunsuke Yabunaka\aff{1}
  \corresp{\email{yabunaka@scphys.kyoto-u.ac.jp}},
  \and 
  Natsuhiko  Yoshinaga\aff{2,3}
  \corresp{\email{yoshinaga@wpi-aimr.tohoku.ac.jp}},
  }

\affiliation
{
\aff{1}
Yukawa Institute for Theoretical Physics, The Kyoto University,
Kitashirakawa Oiwake-Cho, Kyoto, 606-8502, Japan
\aff{2}
WPI - Advanced Institute for Materials Research, Tohoku University,
Sendai 980-8577, Japan
\aff{3}
MathAM-OIL, AIST,
Sendai 980-8577, Japan
}

\maketitle

\begin{abstract}
 We \NY{use analytical and numerical approaches to investigate head-on collisions}
 between two
 self-propelled drops \SSY{described as a phase separated binary mixture}.
 Each drop is driven by chemical reactions that \NY{isotropically produce or consume the
 concentration of a third chemical component, which affects the surface
 tension of the drop}.
The isotropic distribution of the concentration field is destabilized by
 motion of the drop\NY{, which is created by the Marangoni flow from the 
 concentration-dependent surface tension.}
This symmetry-breaking self-propulsion is distinct from other self-propulsion mechanisms 
 due to \NY{its intrinsic polarity of} squirmers and self-phoretic
 motion; there is a bifurcation point below which the drop is stationary
 and above which it moves spontaneously.
When two drops \NY{are moving in the opposite direction along the same axis, their interactions arise from hydrodynamics and concentration
 overlap.}
 We found that two drops exhibit either an elastic collision or fusion,
 depending on the distance from \NY{their bifurcation point, which may
 be controlled, for
 example,} by viscosity.
\NY{An} elastic collision \SSY{occurs} \NY{when there is a} balance between dissipation and
\NY{the injection of} energy by chemical reactions. 
 We derive the reduced equations for the collision between two drops and
 \NY{analyse} the contributions from the two interactions.
 The concentration-mediated interaction is found to dominate
 the hydrodynamic interaction \NY{for a head-on collision}.
\end{abstract}

\section{Introduction}
In biological systems, cells and microorganisms are moving spontaneously
and autonomously by consuming energy from ATP hydrolysis.
The size and swimming speed of \NY{bacteria are small, and their}
self-propulsion is described by \NY{low-Reynolds-number hydrodynamics}.
\NY{In an attempt} to capture the generic mechanism of the self-propulsion,
\NY{various} mathematical models have been investigated.
\NY{One of these is the squirmer model, which considers} the flow
field created by \NY{the}
beating of cilia on \NY{the} body of a microorganism and/or \NY{the
deformation of its body
surface} \citep{Lighthill:1952,blake:1971}.
The problem then \NY{becomes a matter of solving the Stokes equation
under the boundary
conditions of finite velocity on the surface of} the body in the normal
and/or tangential directions.
\NY{The translational and angular velocity of a single squirmer} are well understood \citep{stone:1996}.
Apart from \NY{that} model, there have been several attempts to find other
classes of self-propulsion.
The \NY{simplest} extension of the model is to include an additional scalar field
such as \NY{the} concentration, electric, or temperature field.
\NY{A Janus particle, which is an asymmetric particle with two different
surface properties, creates a gradient of the field around the
particle, and thus, in turn, causes it to move spontaneously} \citep{Paxton:2004,howse:2007,jiang:2010}.
\NY{This} motion is similar to phoresis\NY{, except that the gradient is self-generated
instead of imposed}.
This mechanism is thus called self-phoresis.

The squirmer and Janus particles have intrinsic asymmetry\NY{,} and therefore
the swimming direction is set by the polar direction of the asymmetry \NY{of
each particle}.
This \NY{simplifies the problem:} there is a linear
relation between the self-propulsion speed and the magnitude of the
asymmetry.
\NY{
This idea may also be \SYYYY{extended} toward self-propulsion of a geometrically
asymmetric object with a uniform surface property \citep{Tsemakh:2004,Shklyaev:2014}
}
On the other hand, cells often break \NY{symmetry} to choose the
direction of motion \citep{yam:2007}.
This phenomenon is not captured by the squirmer and Janus particles, and
therefore another class of self-propulsion \NY{must} be considered.
\NY{As a step in} this direction, \NY{various} mathematical models that include the
internal polarity field have been proposed \citep{shao:2010,Ziebert:2012,Tjhung:2012}.
In these models, \NY{it has been observed that spontaneous symmetry
breaking results in directional motion}.

Along this line, it was recently found that \NY{nonliving
chemically driven} systems exhibit self-propulsion \citep{Toyota:2009,Thutupalli2011,Izri:2014}.
In these systems, \NY{a} drop may produce or consume chemical molecules
\NY{in such a way}
that \NY{the system is away from an equilibrium state}.
The flux couples with the motion and results in an asymmetric concentration
distribution.
Once the symmetry is broken, the surface tension becomes anisotropic and
\NY{this} creates flow \NY{both} inside and outside the drop.
\NY{This motion, which occurs along the given gradient of concentration and/or temperature
fields,} is known as the Marangoni effect \citep{Young:1959,Fedosov:1956}.
The self-propulsive motion using the Marangoni effect \NY{resulting from a chemical
reaction was first proposed as a {\it reactive
drop} \citep{ryazantsev:1985}, and later its mechanism was theoretically
reformulated as a bifurcation phenomena}  \citep{Yabunaka:2012,Yoshinaga:2012a,Yoshinaga:2014}.
In these \NY{studies}, the reduced nonlinear equations \NY{were} derived from the
coupled advection-diffusion and hydrodynamic equations.
\NY{A similar idea was considered} for the auto-phoretic motion, \NY{which is
the self-phoretic motion of a Janus particle due to a nonlinear
coupling of an {\it isotropic} chemical reaction and advection} \citep{Michelin:2013}.

\NY{Although the self-propulsion of an isolated particle/drop is well
understood, there is still only a limited understanding of the
interactions between them}.
\NY{There} have been intensive numerical studies of
\NY{interactions} between squirmers \citep{ishikawa:2006} and between \NY{a} squirmer and \NY{a} wall \citep{Spagnolie:2012}.
In particular, \NY{understanding} of the squirmer/wall system has
recently \NY{increased, and numerical simulations have revealed} the bound
state near the wall \citep{Li:2014}.
\NY{These results are consistent with those from an analysis of the
equation of motion of a squirmer, using a technique for analysing dynamical systems} \citep{Ishimoto:2013}.
Even for Janus particles, \NY{numerical
simulations near a wall have only been performed very recently} \citep{Uspal:2014}.

In this work, we discuss the interaction between self-propelled drops.
In particular, we focus on \NY{head-on collisions} between two drops.
As we will discuss, the interaction \NY{arises from hydrodynamics and
a concentration overlap}.
The hydrodynamic interaction has been discussed in \NY{terms of} the
squirmer model\NY{, in which} only the velocity field is treated.
The concentration overlap has been discussed in the \NY{context} of
reaction-diffusion systems\NY{; in that case,} the concentration fields are \NY{analysed}
without considering \NY{the} hydrodynamics, and thus \NY{mechanics does
not play} a role \citep{Ohta:1997a,ohta:2001,Bode:2002,Nishiura:2003,ei:2006}.
The primary questions are which effect dominates the interaction and
when \NY{do crossovers occur}.
To answer \NY{these} questions, we consider the theory for two interacting drops
\NY{that are} separated far away from each other.
This is \NY{an} extension of the theory for a single drop discussed in
\citep{Yabunaka:2012,Yoshinaga:2014}.
\NY{
Other studies \citep{Golovin:1995,Lavrenteva:1999} have used a
boundary-value approach to investigate the
interaction that arises from hydrodynamics and the concentration (or heat)
field.
Our model shares a similar philosophy, although we focus on
the equations of motion of a reduced description and drops with unsteady
motion, rather than stationary speed.
The main difference is that we use a diffuse-interface approach, and
because of this, both analytical and numerical solutions become tractable.
We will discuss the similarities and differences in section \ref{sec.summary}.
}

We also develop \NY{numerical simulations} of isolated as well as
interacting drops.
This enables us to investigate the effect of advection of the chemical
component; a complete analytical investigation of this has not been
previously performed \citep{Yabunaka:2012, Yoshinaga:2014}.
We confirm that the convection of the chemical component does not change
\NY{the} essential bifurcation of a single drop\NY{, but it} suppresses the drift
instability\NY{; this supports \SSY{previously} presented theories} \citep{Yabunaka:2012,
Yoshinaga:2014}.
For interacting drops, we will numerically \NY{investigate the dynamics
of collisions; this complements our theoretical calculations}.

The interaction between two self-propelled particles is distinct from
\NY{that seen in} conventional passive systems\NY{,} where particles and drops are driven by
external forces \citep{Jeffrey:1984}.
The dominant hydrodynamic interaction \NY{in the} far field does not arise from
\NY{a} Stokeslet but from \NY{a} source doublet or \NY{a} stresslet
depending on the \NY{mode ($l=1$ or $l=2$)} for the expansion of the
slip velocity\NY{, as expressed by} spherical harmonics \citep{Lauga2009,Pak:2014a}.
In addition, our system is different from \NY{either} the squirmer
\NY{or} the Janus particle; \NY{it} does not have \NY{a specific
intrinsic polarity, but polarity
spontaneously appears when the bifurcation parameter exceeds a threshold
value}.
Consequently, \NY{the direction of motion of a drop may change} without rotation.

In the remaining sections, we formulate \NY{a} model for \NY{chemically driven
self-propulsion that uses the} Marangoni effect.
\NY{To prepare for the main parts}, in section \ref{sec.marangoni}, we
compute the flow field and resulting velocity of \NY{a} drop under \NY{a given
distribution of the surface tension}.
In section \ref{sec.isolated}, \NY{we summarize the spontaneous motion of an isolated drop}.
In section \ref{sec.interaction}, \NY{we derive the hydrodynamic and
concentration-mediated interactions between two drops, and the equations of motion for two
interacting drops} are formulated in section \ref{sec.amp.eq}.
\NY{Numerical} results for isolated and colliding drops are \NY{presented} in
section \ref{sec.numerical.isolated} and section
\ref{sec.numerical.interaction}, respectively.
We compare the \NY{numerical} results with our theoretical analysis \NY{of} section \ref{sec.interaction}.
We conclude with section \ref{sec.summary}\NY{, which summarizes} our results.

\section{An isolated drop under a given concentration gradient}
\label{sec.marangoni}

Before discussing the interaction between spherical drops, we first
calculate the flow field around \NY{a} spherical drop driven by
\NY{an arbitrary distribution of the surface tension}.
\NY{The} axisymmetric flow field around the drop has been
studied \citep{Young:1959,levan:1981,Kitahata:2011}.  
Here we do not assume \NY{that the system is axisymmetric; thus, instead
of expanding the surface tension in terms of Legendre polynomials, we use spherical harmonics:}
\begin{equation}
\gamma (\theta, \varphi)
=
\sum_{l,m} \gamma_{lm} Y_l^m (\theta,\varphi)
.
\end{equation}
The velocity fields inside and outside the drop are \NY{expanded as follows:}
\begin{eqnarray}
{\bf v}^{(i)} 
 &=
 \sum_{l,m}
 {\bf v}^{(i)}_{lm}
 =
\sum_{l,m} \left[
f_{lm}^{(i)}(r) {\bf Y}_{lm} (\theta,\varphi)
+  g_{lm}^{(i)} (r) \boldsymbol{\Psi}_{lm} (\theta,\varphi)
\right]
\\
{\bf v}^{(o)} 
 &=
  \sum_{l,m}
 {\bf v}^{(o)}_{lm}
 =
\sum_{l,m} \left[
f_{lm}^{(o)}(r) {\bf Y}_{lm} (\theta,\varphi)
+  g_{lm}^{(o)} (r) \boldsymbol{\Psi}_{lm} (\theta,\varphi)
\right]
,
\end{eqnarray}
\NY{where the outer and inner fields are indicated by the superscripts (o) and (i),
respectively, and the vector spherical harmonics are defined by using the scalar
spherical harmonics $Y_l^m(\theta,\varphi)$, as follows:}
\begin{align}
 \mathbi{Y}_{lm}
&=
Y_l^m \hat{{\bf r}}
\\
 \mathbi{\Psi}_{lm}
&=
r \nabla Y_l^m
,
\end{align}
where $\hat{{\bf r}}$ is a unit normal vector.
Since the flow field is driven by \NY{the} gradient of \NY{the} surface tension, one of
the vector spherical harmonics, $\mathbi{\Phi}_{lm} = {\bf r} \times \nabla
Y_l^m$\NY{, which is} in the tangential
direction perpendicular to $\mathbi{\Psi}_{lm}$\NY{,} does not appear in this expansion.
\NY{Note that} $f_{lm}(r)$ and $g_{lm}(r)$ are determined from the boundary conditions\NY{,} as discussed below.

We consider \NY{a} spherical  drop \SSY{that} is moving with velocity ${\bf u}$
in \NY{an} arbitrary direction.
\NY{At any point $(\theta,\varphi)$ on the drop
surface, the velocity is expressed as}
\begin{align}
{\bf u}
&=
\sum_{m} u_m
\left(
{\bf Y}_{1,m} (\theta,\varphi)
 + \mathbi{\Psi}_{1,m} (\theta,\varphi)
\right)
,
\end{align}
\NY{which can be expressed in the Cartesian coordinates as  (see (\ref{BC.vn}))}
\begin{align}
{\bf u}
&=
\left(
\sqrt{\frac{3}{4 \pi}}
\frac{u_{-1} - u_{1}}{\sqrt{2}}
,
i \sqrt{\frac{3}{4 \pi}}
\frac{-u_{-1} - u_{1}}{\sqrt{2}}
,
\sqrt{\frac{3}{4\pi}} 
u_0
\right).
\end{align}

We \NY{will assume the velocity of the drop is sufficiently slow that the 
Reynolds number is near zero, and thus} it satisfies the Stokes equation except over the surface of the drop $r=R$
\begin{align}
{\eta}^{(i)} \nabla^2 {\bf {v}}^{(i)} -\nabla p^{(i)}
&=0,
\label{stokes.eq}
\\
 {\eta}^{(o)} \nabla^2 {\bf {v}}^{(o)} -\nabla p^{(o)}
&=0.
\label{stokes.eq2}
\end{align}
The pressure $p$ is determined from the
incompressibility condition
\begin{align}
\nabla \cdot { \bf v} 
&= 0.
\end{align}
The boundary conditions on $r=R$ \NY{are}
\begin{align}
 {\bf v}^{(o)} \cdot \hat{\bf r} 
&=  {\bf v}^{(i)} \cdot \hat{\bf r} 
= {\bf u} \cdot \hat{\bf r}
\label{BC.vn}
\\
{\bf v}^{(o)} \cdot \hat{\bf t} 
&=  {\bf v}^{(i)} \cdot \hat{\bf t} 
\label{BC.vt}
\\
\sigma_{nt}^{(i)}(R) \hat{\bf t}
&=
\sigma_{nt}^{(o)}(R) \hat{\bf t}
+ \frac{1}{R} \nabla_s \gamma
\label{BC.stress}
\end{align}
\NY{where $\hat{\bf t}$ is the unit tangent vectors.
The conditions} (\ref{BC.vn}) and (\ref{BC.vt}) are
continuity of the
velocity field across the interface, and (\ref{BC.stress}) \NY{implies
that the forces are balanced at the interface, that is,} the jump of shear stress across
the interface due to the force created by the inhomogeneous surface
tension \citep{Scriven:1960}.
The system is force-free; there is no mechanical force acting on the drop, 
\begin{align}
{\bf F} 
&=
\int dS \sigma^{(o)}(R) \cdot {\bf n}
=0,
\end{align}
where the integral is taken over the surface of the drop.
The stress balance in the normal direction is \NY{automatically} satisfied
for the $l=1$ mode and determines \NY{the} shape of the drop for $l \geq 2$.

The solution of the Stokes equations \NY{for this system,
(\ref{stokes.eq}) and (\ref{stokes.eq2}), can be decomposed into
two parts: one for $l=1$ and the other for $l \geq 2$}.
For $l=1$,
\begin{align}
{\bf v}^{(i)}_{1,m} 
&=
u_m
\left(
\left[
-\frac{3}{2}
\left( \frac{r}{R} \right)^2
+ \frac{5}{2}
\right] {\bf Y}_{1,m}
+
\left[
-3
\left( \frac{r}{R} \right)^2
+ \frac{5}{2}
\right] \mathbi{\Psi}_{1,m}
\right)
\label{vel.isolated.in.1}
\\
{\bf v}^{(o)}_{1,m} 
&=
u_m
\left[
\left( \frac{R}{r} \right)^3
 {\bf Y}_{1,m}
-\frac{1}{2}
\left( \frac{R}{r} \right)^3
 \mathbi{\Psi}_{1,m}
\right]
\label{vel.isolated.out.1}
\\
p^{(o)}
&=0
\\
p^{(i)}
&= -
\sum_{m}
\frac{10 \eta^{(i)} \gamma_{1,m}}{R (3 \eta^{(i)} + 2 \eta^{(o)})}
\frac{r}{R} Y_1^m (\theta,\varphi)
= -
\sum_{m}
\frac{15 \eta^{(i)} u_m}{R }
\frac{r}{R} Y_1^m (\theta,\varphi)
\end{align}
and for $l \geq 2$
\begin{align}
{\bf v}^{(i)}_{l,m} 
=&
\frac{\gamma_{lm}}{2 (\eta^{(i)} + \eta^{(o)}) (2l+1)}
\left(
l(l+1)
\left[
\left( \frac{r}{R} \right)^{l+1}
- \left( \frac{r}{R} \right)^{l-1}
\right] {\bf Y}_{lm}
\right.
\nonumber \\
& \left.
+
\left[
-(l+1)
\left( \frac{r}{R} \right)^{l-1}
+ (l+3) \left( \frac{r}{R} \right)^{l+1}
\right] \mathbi{\Psi}_{lm}
\right)
\label{vel.isolated.in.l}
\\
{\bf v}^{(o)}_{l,m} 
= &
\frac{\gamma_{lm}}{2 (\eta^{(i)} + \eta^{(o)}) (2l+1)}
\left(
l(l+1)
\left[
\left( \frac{R}{r} \right)^l
- \left( \frac{R}{r} \right)^{l+2}
\right]
 {\bf Y}_{lm}
\right.
\nonumber \\
& \left.
+
\left[
- (l-2)
\left( \frac{R}{r} \right)^l
+ 
l \left( \frac{R}{r} \right)^{l+2}
\right]  \mathbi{\Psi}_{lm}
\right)
\label{vel.isolated.out.l}
\\
p^{(o)}
&=
\sum_{l,m}
\frac{\eta^{(o)} \gamma_{lm}}{R (\eta^{(i)} + \eta^{(o)})}
\frac{
l
\left(
2l-1
\right)
}{2l+1}
\left(
\frac{R}{r}
\right)^{l+1}
Y_l^m (\theta,\varphi)
\\
p^{(i)}
&=
\sum_{l,m}
\frac{\eta^{(i)} \gamma_{lm}}{R (\eta^{(i)} + \eta^{(o)})}
\frac{
(l+1)
\left(
2l + 3
\right)
}{2l+1}
\left(
\frac{r}{R}
\right)^{l}
Y_l^m (\theta,\varphi)
.
\end{align}
Because of the force-free condition, the velocity field for $l=1$ decays
as $1/r^3$ and not as $1/r$.
This \NY{occurs because} the motion is not driven by
the Stokeslet but \NY{by} the quadrupole (source dipole).
The velocity of the drop is
\begin{align}
u_m 
&=
- \frac{2 \gamma_{1,m}}{ 3 (3 \eta^{(i)} + 2 \eta^{(o)})}
.
\end{align}
The axisymmetric case corresponds to $\gamma_{l,m} = 0$ for $m \neq 0$.
\NY{This} result is consistent with \NY{that of} \citep{Young:1959} 
\footnote{
except \NY{for typographical errors;} See \citep{levan:1981,Kitahata:2011}
}.
Note that the coefficient may depend on the definition of the
normalization factor in the spherical harmonics.

\begin{figure}
\begin{center}
\includegraphics[width=0.95\textwidth]{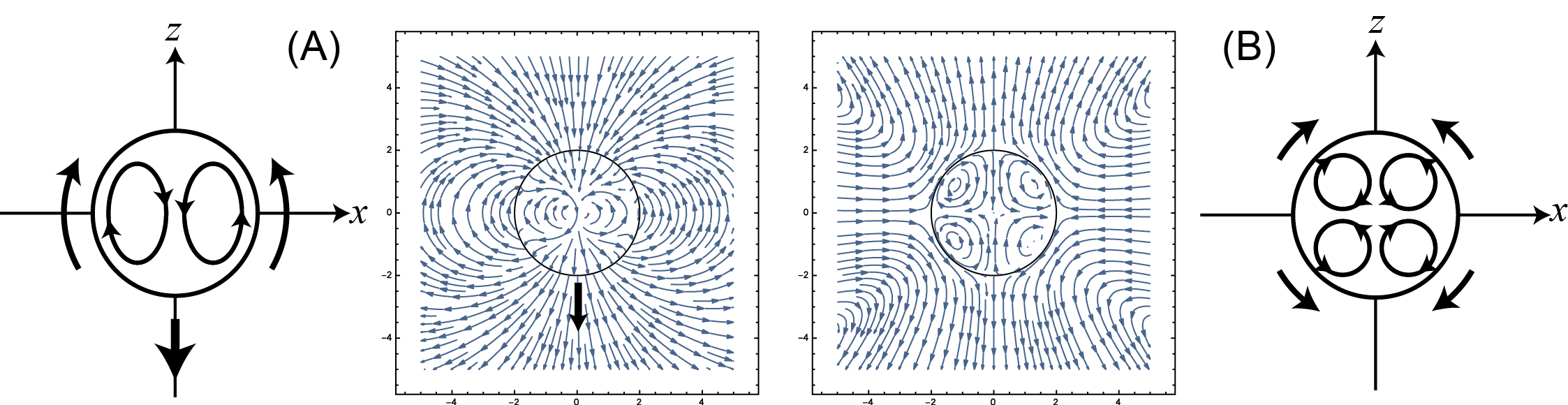}
\caption{ 
The flow field created by \NY{an} isolated drop. (A) The $l=1$ mode
 associated with the translational motion under $\gamma_{1,0}>0$\NY{, as}
 given
 by (\ref{vel.isolated.in.1}) and (\ref{vel.isolated.out.1}).  
The direction of motion is \NY{indicated by} the thick arrow.
(B) The $l=2$ mode of the
 dipolar flow for $\gamma_{2,0} >0$\NY{, as} given by
 (\ref{vel.isolated.in.l}) and (\ref{vel.isolated.out.l}).
 The schematic drawings of the flow fields are also shown.
 \label{fig.flow.schematic}
}
\end{center}
\end{figure}


\section{Self-propelled motion of a single drop}
\label{sec.isolated}

In this section, we summarize the \NY{self-propelled} motion of an isolated
drop.
The details for the analysis of this model \NY{can be} found in
\citep{Yabunaka:2012}.
We consider the dynamics of the concentration field\NY{, $c({\bf r})$,
of a third dilute component} 
\begin{align}
\pdiff{c}{t} 
+ {\bf v} \cdot \nabla c
&=
D\nabla^2 c - \kappa (c-c_{\infty})
+ 
A \Theta \left(
R - |{\bf r} - {\bf r}_{G}|
\right)
.
\label{con}
\end{align}
The system is driven away from \NY{equilibrium}, where the drop is
stationary, by the source term described by the step function
$\Theta(x)$ with the coefficient $A$.
When $A>0$, the drop produces chemical molecules, while when $A<0$, the
drop consumes them.
Note that we do not use the \NY{comoving frame with the drop;
therefore, the advection term in (\ref{con}) describes only the
advection due to}
the flow.
 \NY{In a comoving frame, there will be an additional contribution due
 to} the
 fact that the drop is moving.
In our model, this effect is included in the last term in (\ref{con}).

\NY{Based on} the diffuse-interface model \citep{anderson:1998,
hohenberg:1977}, we describe the drop as a binary mixture\NY{, where} $\phi =1$ inside the drop and
$\phi =-1$ outside.
The dynamics is given by \NY{the} Cahn-Hilliard equation with advection\NY{:}
\begin{align}
\pdiff{\phi}{t} + {\bf v} \cdot \nabla \phi
&=
\nabla \cdot L \nabla \frac{\delta F}{\delta \phi}
\label{DphiDt}
\end{align}
where the mobility $L$ is assumed to be constant.
The free energy \NY{has a} double-well type potential\NY{:}
\begin{align}
F 
&=
\int_{\Omega}
\left[
-\frac{1}{2} \phi^2 + \frac{1}{4} \phi^4
+ \frac{B(c)}{2} \left| \nabla \phi \right|^2
\right]
\end{align}
where $\Omega$ is \NY{the entire} domain of the system.
The interfacial energy is dependent on the concentration field $c({\bf r})$.
We assume the linear relation 
\begin{align}
B(c) 
&=
B_0 + B_1 c({\bf r}),
\label{B.linear.c}
\end{align}
\NY{which is} characterized by two parameters\NY{,} $B_0$ and $B_1$.
The benefit of this approach is that we do not need to solve the Stokes
equation with moving boundary conditions.
The force acting on the fluid is given by
\begin{align}
{\bf f} 
&=
- \phi \nabla  \frac{\delta F}{\delta \phi}
- c \nabla  \frac{\delta F}{\delta c}
;
\end{align}
\NY{this is} generated by the surface tension and thus \NY{localises}  at
the interface between the drop and \NY{the} surrounding fluid.
The force \NY{can be} expressed in \NY{divergence form as}
\begin{align}
{\bf f} 
&=
\nabla \cdot \Pi
\end{align}
where the stress is
\begin{align}
\Pi_{ij}
&=
B(c) \nabla_i \phi \nabla_j \phi 
\mbox{
+ isotropic terms
}
.
\end{align}
The isotropic terms merely \NY{modify} the reference pressure and thus
we \NY{will not discuss them further.}
\NY{
In the thin-interface limit, the concentration in the region of
the diffused interface (in bulk) can be represented by the surface
concentration.
Then, (\ref{B.linear.c}), expressing the gradient energy at a 
given bulk concentration, generally leads to a nonlinear dependence of
the surface tension on the surface concentration.
However,  the surface tension can be linearly expanded with respect to
the deviation from  a constant value $c_0$ of the surface concentration,
provided that the deviation across the entire surface is small:
\begin{align}
\gamma(c_I) 
&=
\gamma(c_0) + \left. \frac{\partial \gamma(c_0)}{\partial c_I} (c_I-c_0) \right|_{{\bf r} = {\bf R}}
\label{gamma}
,
\end{align}
where $c_I$ represents the surface concentration.
We may also be able to consider a more realistic dependence by replacing
the functional form in (\ref{B.linear.c}) with a logarithmic function.
}

Instead of (\ref{stokes.eq}) and (\ref{stokes.eq2}), we solve the
\NY{following} single Stokes equation \NY{with the force ${\bf f}$ for the entire space under the incompressibility condition $ \nabla \cdot { \bf
v} = 0$:}
\begin{align}
\eta \nabla^2 {\bf v} -\nabla p 
+ {\bf f}
 &=0
 .
\label{stokes.force.eq}
\end{align} 
We also assume $\eta^{(o)}=\eta^{(i)}$.

As discussed in \citep{Yabunaka:2012}, \NY{this} model leads to the following reduced description:
\begin{align}
m \odiff{{\bf u}}{t}
&=
\left(
- \tau_c + \tau
\right) {\bf u} - g {\bf u} |{\bf u}|^2
\label{amp.eq.single}
.
\end{align}
In deriving this equation, they assumed \NY{the following:} (i)  the migration of the
drop due to the diffusion is much slower than that due to the flow field, (ii) the contribution due to the deformation of the drop is
negligible and (iii) the convective term in (\ref{con}) does not
\NY{qualitatively affect the bifurcation}.
\NY{Assumption} (i) can be justified when $R \gg (L \eta_0)^{1/2}$ by
estimating the migration velocity due to the gradient of the
concentration $c$ \citep{Yabunaka:2012, Bhagavatula:1997}.
\NY{Assumption} (ii) is
justified when $\gamma_c A/(\gamma \kappa) \ll 1$ \citep{Yoshinaga:2014}.

Here the velocity and time are rescaled as \NY{follows:} 
\begin{align}
\frac{{\bf u} }{D \beta}
&\rightarrow
{\bf u}
\\
 \kappa t 
&\rightarrow
t
.
\end{align}
The length \NY{is} also rescaled by \NY{the} inverse length $\beta$\NY{,
which is defined as follows:}
\begin{align}
\beta 
&=
\sqrt{\frac{\kappa}{D}}
.
\end{align}
The coefficients $m$, $\tau$, and $g$ are dependent only on $\beta R$.
\NY{Note that} (\ref{amp.eq.single}) is characterized by the single parameter 
\begin{align}
\tau_c 
&=
\frac{15 \eta D^2 \beta^3}{2 \gamma_c A}
\label{tauC}
.
\end{align}
The drop is stationary \SYYYY{for} $\tau < \tau_c$,
\SYYYY{which occurs when the rate of the chemical reaction is small  ($A \ll 1$) or the
viscosity is high ($\eta \gg 1$).
At $\tau = \tau_c$,
}
\NY{bifurcation occurs,} and the drop starts to move.
Solving (\ref{amp.eq.single}), the speed of the drop is
\begin{align}
u 
&= |{\bf u}|
=
\begin{cases}
0
\mbox{   for   }
\tau < \tau_c
\\
\frac{u_0 e^{t/s_r}}{ \sqrt{1+ \left( e^{2t/s_r} - 1\right)
 \frac{u_0^2}{u_{st}^2}}}
\mbox{   for   }
\tau \geq \tau_c,
\label{steady.vel.single}
\end{cases}
\end{align}
where the relaxation time $s_r$ and the steady velocity $u_{st}$ are
\begin{align}
s_r 
&=
\frac{m}{\tau - \tau_c}
\\
u_{st}
&=
\sqrt{\frac{m}{g s_r}}
\label{steady.vel.single2}
.
\end{align}
\NY{Note} that, as the nondimensional reaction rate $\tau_c$ \NY{approaches} $\tau$, the relaxation time $s_r$ diverges.
\NY{It can be shown that, near the bifurcation point, the surface concentration
deviation is very small across the entire region of the diffused interface; this justifies the assumption made for (\ref{gamma}).}
\section{Interacting Drops}
\label{sec.interaction}

When $N$ drops are placed at disconnected positions, the concentration field is described by
\begin{align}
\pdiff{c}{t} 
+ {\bf v} \cdot \nabla c
&=
D\nabla^2 c - \kappa (c-c_{\infty})
+ 
\sum_{i=1}^{N}
A_i \Theta \left(
R_i - |{\bf r} - {\bf r}_{G,i}|
\right)
\label{con.N}
\end{align}
where \NY{$A_i$, $R_i$, and ${\bf r}_{G,i}$ are the source strength,
size, and centre of mass of the $i$th drop, and $c_{\infty}$ is the concentration at infinity}.
Here we consider $N=2$ drops \NY{of the same mean size of drops, $R_1=R_2 =
R_0$, and} we set $c_{\infty}=0$.
We assume \NY{a} sufficiently large bare surface tension $\gamma_0$\NY{,} so that the shape of
the drop is always spherical.
\NY{We} neglect the advection term ${\bf v} \cdot \nabla c$ in
(\ref{con.N}) \NY{until section \ref{sec.numerical.isolated}, where its
effect will be discussed.}

\begin{figure}
\begin{center}
\includegraphics[width=0.40\textwidth]{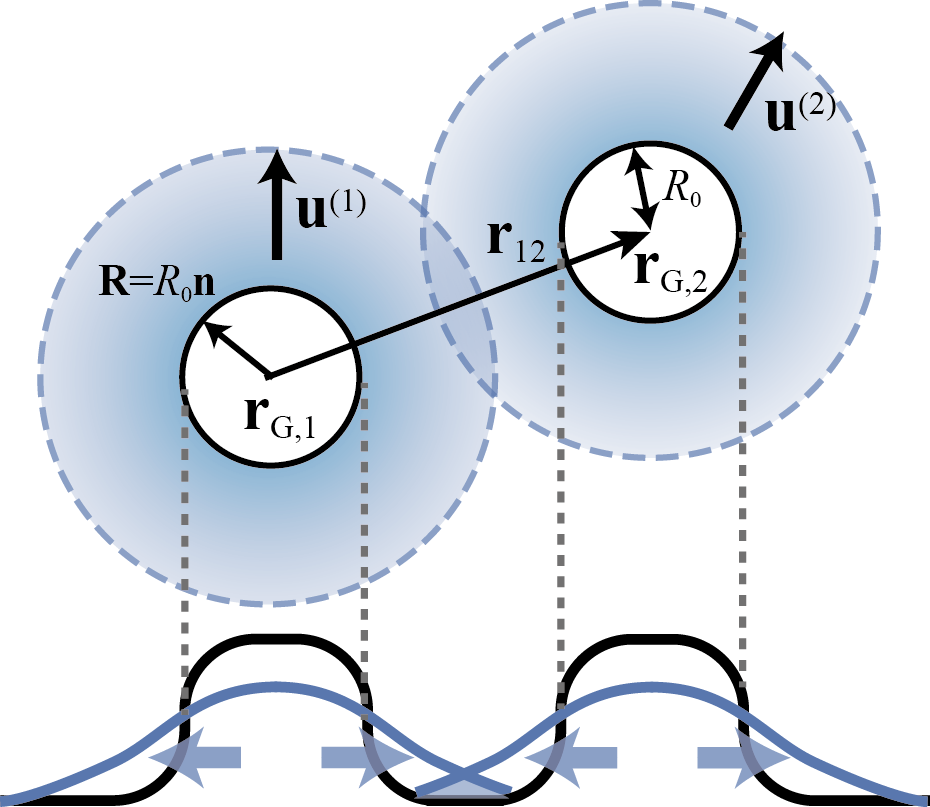}
\caption{
 A schematic \NY{representation} of interacting spherical drops.
 \NY{Each drop produces an outer concentration field}.
 The black line shows the
 drop described by the field $\phi ({\bf r})$.
 The blue (grey) lines \NY{indicate the
 concentration fields that are} independently created by each drop.
 The total
 concentration field $c({\bf r})$ contains \NY{the overlap between the
 two fields, and the two drops interact through this field}.
 (bottom) \NY{Side} view of
 the fields $\phi ({\bf r})$ and $c({\bf r})$.
 \label{fig.schematics}
}
\end{center}
\end{figure}

The velocity of \NY{each} drop is given by \citep{Kawasaki1983,Yabunaka:2012}
\begin{align}
{\bf u}
 &=
 \frac{1}{V} \int  v_n {\bf R} dS
 \label{velocity.drop.def}
\end{align}
\NY{where} $V = \frac{4}{3} \pi R_0^3$.
\NY{Note that the velocity is different 
from that for an isolated self-propelled drop; this is due to
 the concentration field  ${\bf u}_c$ (see Figure~\ref{fig.schematics}) and
the hydrodynamic interaction ${\bf u}_h$}.
The velocity \NY{can be} expressed as
\begin{align}
{\bf u} 
 &=
 {\bf u}_0
+ {\bf u}_c
+ {\bf u}_{h},
\label{vel.sum}
\end{align}
\NY{
where ${\bf u}_0$ is the contribution from an isolated drop.
In the following sections, we will compute ${\bf u}_{h}$
(section~\ref{sec.hydro.int}) and ${\bf u}_0 + {\bf u}_c $ (section~\ref{sec.conc.int}).
}

\subsection{Hydrodynamic interaction}
\label{sec.hydro.int}

First, we consider the hydrodynamic interaction.
By assuming \NY{that} the two drops \NY{are far away} from each other, we may replace the contribution of the hydrodynamic interaction ${\bf u}_h$
in (\ref{vel.sum})
by the flow field created by the second drop:
Using Faxen's law \citep{Hetsroni:1970},
the hydrodynamic interaction \NY{can be} expressed in terms of the flow
field $ {\bf v}^{(2)}$ generated by the second drop, as \NY{follows:}
\begin{align}
 {\bf u}_{h}
&=
\left. {\bf v}^{(2)} \right|_{{\bf r} = {\bf r}_1} 
+ 
\mathcal{O} \left( 
\left. \nabla^2 {\bf v}^{(2)} \right|_{{\bf r} =
 {\bf r}_1} 
\right)
.
\end{align}
The velocity is evaluated at the \NY{centre} of the first drop.
Because of the Laplacian operating on the velocity field, the second
term is negligible compared with the first term when the distance
between \NY{the} two drops is large\NY{, that is,} $r_{12} \gg R$.
\NY{Near} the drift bifurcation point \NY{(distance $\epsilon$)},
the velocity of the second drop \NY{ is as $u^{(2)} \sim \epsilon$, and
the surface tension scales as}
$\gamma_{l,m} \sim \epsilon^l$.
\NY{The} velocity field decays \NY{as} $1/r^3$ for $l=1$ and \NY{as} $1/r^2$ for
$l=2$.
Therefore, it suffices to consider $l=1$ and $l=2$.
The flow created by the second drop is
\begin{align}
{\bf v}^{(1)}_{1,m} 
& \simeq 
u_m^{(2)}
\left[
\left( \frac{R_0}{r_{12}} \right)^3
 {\bf Y}_{1,m} (\pi - \theta_{12}, \pi + \varphi_{12})
-\frac{1}{2}
\left( \frac{R_0}{r_{12}} \right)^3
 \mathbi{\Psi}_{1,m} (\pi - \theta_{12}, \pi + \varphi_{12})
\right]
\\
{\bf v}^{(1)}_{2,m} 
& \simeq
\frac{3 \gamma_{2,m}^{(2)}}{5 (\eta^{(i)} + \eta^{(o)})}
\left( \frac{R_0}{r_{12}} \right)^2
 {\bf Y}_{2,m} (\pi - \theta_{12}, \pi + \varphi_{12})
,
\end{align}
where $\theta_{12}$ and $\varphi_{12}$ are the polar and azimuthal angles of ${\bf r}_2- {\bf r}_1$, respectively. 
\NY{
Here, ${\bf v}^{(1)}_{1,m}$ is the quadrupole flow created by the second drop
perturbing the first drop.
This flow decays as $1/r_{12}^3$.
The dipolar flow generated by the second drop is ${\bf
v}^{(1)}_{2,m}$, which decays as $1/r_{12}^2$.
It should be noted that unlike squirmer and Janus particles, the
far-field flow is not necessarily dominated by the dipolar flow.
This is because the second mode of the surface tension
$\gamma_{2,m}^{(2)}$ associated with the ellipsoidal concentration
field becomes small near the critical point of the drift bifurcation \citep{Yoshinaga:2014}.
}

\subsection{Concentration-mediated interaction}
\label{sec.conc.int}

Next, we consider \NY{the} interaction between two drops \NY{due to} overlap of the
concentration field.
 \NY{We follow} the approach in \citep{Ohta:1997a,ohta:2001} (\NY{see} Figure \ref{fig.schematics}).
In Fourier space, (\ref{con.N}) \NY{is}
\begin{align}
\pdiff{c_{\bf q}}{t}
&=
-D (q^2 + \beta^2) c_{\bf q}
+ H_{\bf q},
\label{con.fourier.pair}
\end{align}
\NY{
where the source term $H_{\bf q}$ is
 }
\begin{align}
H_{\bf q} 
&=
A_1
S_q^{(1)}  e^{i {\bf q} \cdot {\bf r}_{G,1}}
+ A_2
 S_q^{(2)} e^{i {\bf q} \cdot {\bf r}_{G,2}}
 \label{Hq1and2}
\end{align}
and
\begin{align}
S^{(1)}_q 
&=
S^{(2)}_q 
=S_q 
=
4 \pi \frac{\sin (q R_0) - q R_0 \cos (qR_0)}{q^3}
= \frac{4\pi R_0^2}{q} j_1(q R_0)
.
\end{align}
\NY{
The first term in (\ref{Hq1and2}) corresponds to the production of
chemicals from the first drop (when $A_1>0$) while the second term corresponds to
production from the second drop (when $A_2>0$).
}
Here $j_n(x)$ for $n=0,1,2,....$ are spherical Bessel functions\NY{, as} defined in (\ref{sphbessel}).
\NY{As in} \citep{Yabunaka:2012}, the solution of (\ref{con.fourier.pair}) is expanded close to the critical
point of \NY{the} drift bifurcation, \NY{that is,} for $\epsilon = u/(D\beta) \ll 1$,
\begin{align}
c_{\bf q} 
&=
\frac{G_q}{D} H_{\bf q}
- \frac{G_q^2}{D^2} \pdiff{H_{\bf q}}{t}
+ \frac{G_q^3}{D^3} \pdiff{^2 H_{\bf q}}{t^2}
- \frac{G_q^4}{D^4} \pdiff{^3 H_{\bf q}}{t^3}
+ \cdots
\label{con.sol.fourier.pari}
\end{align}
\NY{where we use} the Green's function
\begin{align}
G_q 
&=
\frac{1}{q^2 + \beta^2}.
\end{align}
\NY{
Note that the time derivative of $H_{\bf q}$ generate the velocity of the first or second drop (and their time derivative).
After performing inverse Fourier transformation
of (\ref{con.sol.fourier.pari}), the concentration $c_I$ at the
 interface of the first drop 
 can be expanded as
\begin{align}
c_I 
&=
c_I^{(0)} ({\bf r}_{G,1} + {\bf s}) 
+ c_I^{(1)} ({\bf r}_{G,1} + {\bf s}) 
+ c_I^{(2)} ({\bf r}_{G,1} + {\bf s}) 
+ c_I^{(3)} ({\bf r}_{G,1} + {\bf s})
 + \cdots
 .
\label{cI.pair}
\end{align}
The lowest-order term in (\ref{cI.pair}) can be explicitly written as 
}
\begin{align}
c_I^{(0)} ({\bf r}_{G,1} + {\bf s}) 
&=
\frac{1}{D} \int_{\bf q}
G_q 
\left[
A_1 S_q^{(1)}
e^{i {\bf q}\cdot {\bf r}_{G,1}}
+ 
A_2 S_q^{(2)}
e^{i {\bf q}\cdot {\bf r}_{G,2}}
\right]
e^{-i {\bf q}\cdot ({\bf r}_{G,1}+{\bf s})}
\nonumber \\
&=
\frac{A_1}{D} 
\left[
Q^{(0)}_1(s)
+ Q^{{\rm int}}_1(\theta,\varphi)
\right]
,
\label{cI0.pair} 
\end{align}
\NY{
where the terms correspond to the respective terms in
(\ref{Hq1and2}); the first term arises from self-production of the
chemical concentration while the
second term is from the interaction.
The first term $Q_n^{(0)} (s)$ is
}
\begin{align}
 Q_n^{(0)} (s) 
&=
\int_{\bf q} G_q^n S_q e^{-i{\bf q}\cdot {\bf s}}
\nonumber \\
&=
\frac{2 {R_0}^2 }{\pi}
\int_0^{\infty}  dq 
G_q^n q j_1(q R_0)
j_0 (q s) 
\label{Qn0.pair}
.
\end{align}
\NY{
There is no angular dependence, and thus this term describes an isotropic
concentration field.
Without hydrodynamic flow and the resulting motion of the drops, our model
is isotropic, and therefore, the lowest-order concentration field must be
isotropic.
Nevertheless, as we will see, the coupling to the flow field or perturbation of the
concentration field by another drop would result in an anisotropic
concentration field, which would then lead to an inhomogeneous surface tension and self-propulsion.  
}
The contribution from interaction, 
$Q^{{\rm int}}_n (\theta,\varphi)$
\NY{
in (\ref{cI0.pair})
can be calculated as
}
 \begin{align}
  Q_n^{{\rm int}} ({\bf s})
&=
\frac{A_2}{A_1}
\int_{\bf q} G_q S_q 
e^{-i{\bf q}\cdot \left( {\bf s}+ {\bf r}_{G,1} - {\bf r}_{G,2} \right)}
\nonumber \\
&=
\frac{2  {R_0}^2 A_2}{\pi A_1}
\int_0^{\infty} dq 
G_q^n q
j_0 \left( 
q|{\bf s}+ {\bf r}_{G,1} - {\bf r}_{G,2}|
\right)
j_{1} (q R_0)
,
\label{Qn02}
 \end{align}
Using the addition theorem of \NY{spherical Bessel functions}
\citep{watson:1922}, we have
\begin{align}
j_0 \left( 
q|{\bf s}+ {\bf r}_{G,1} - {\bf r}_{G,2}|)
\right) 
&=
\sum_{l=0}^{\infty}
(2l+1)
j_l (qs)
j_l (q r_{12})
P_l (\cos \phi_{s12}),
\label{Legendre.addition}
\end{align}
where
$r_{12} = |{\bf r}_{G,2} - {\bf r}_{G,1}|$ and $\phi_{s12}$ is the
angle between ${\bf s}$ and ${\bf r}_{12} = {\bf r}_{G,2} - {\bf r}_{G,1}$.
The Legendre polynomial \NY{can be decomposed as follows \citep{Arfken:1968}:}
\begin{align}
P_l (\cos \phi_{s12}) 
&=
\frac{4\pi}{2l+1}
\sum_{m=-l}^{l}
Y_l^m (\theta,\varphi) 
 Y_l^{m*} (\theta_{12},\varphi_{12})
 .
\label{SphericalHarmonics.addition}
\end{align}
Then (\ref{Qn02}) becomes
\begin{align}
 Q_n^{{\rm int}} ({\bf s})
&=
\frac{8  {R_0}^2 A_2}{A_1}
\sum_{l=0}^{\infty} \sum_{m=-l}^l
\int_0^{\infty} dq 
G_q^n q
j_{1} (q R_0)
j_l (qs)
j_l (q r_{12})
Y_l^m (\theta,\varphi) 
Y_l^{m*} (\theta_{12},\varphi_{12}) 
.
\label{Qn02b}
\end{align}
\NY{
This concentration field, (\ref{cI0.pair}) and (\ref{Qn02b}), becomes
anisotropic, since it contains $Y_l^m (\theta,\varphi)$.
This arises from the coupling of the relative position $Y_l^{m*} (\theta_{12},\varphi_{12})$ to the concentration field created by
another drop in contrast with the isotropic term of (\ref{Qn02}).
}

\NY{
In expansion of the concentration field, (\ref{cI.pair}), the next order
term is
}
\begin{align}
c_I^{(1)} ({\bf r}_G + {\bf s}) 
&=
- \frac{A_1}{D^2}
\int_{\bf q} G_{q}^2
\left[
(i {\bf q} \cdot {\bf u}^{(1)})
S_q^{(1)}
e^{i{\bf q} \cdot {\bf r}_{G,1} }
+
\frac{A_2}{A_1}
(i {\bf q} \cdot {\bf u}^{(2)})
S_q^{(2)}
e^{i{\bf q} \cdot {\bf r}_{G,2} }
\right]
e^{-i{\bf q} \cdot ({\bf r}_G + {\bf s})}
\nonumber \\
&=
u_i^{(1)} \frac{A_1}{D^2}
\left[
n_i^{(0)}
\pdiff{Q_2^{(0)}}{s}
\right]
+ 
u_i^{(2)} \frac{A_2}{D^2}
\pdiff{}{s_i}
 Q_2^{{\rm int}}\SYYYY{,}
 \label{cI1}
\end{align}
\NY{
where the first term arises from the source term of the first drop and thus is the same as the velocity without the
second drop.
In contrast with the lowest-order concentration $c_I^{(0)}$, both of the
terms in (\ref{cI1}) are anisotropic.
In the first term, this is because the coupling to the velocity of the
drop (${\bf u}^{(1)}$) and the concentration field.
Since the drop is moving, the produced concentration field remains at
the back of the drop, leading different concentrations between at the front and rear \citep{Yabunaka:2012}. 
In the second term, both the velocity of the drop and the interaction
produce an anisotropic concentration.
}

\begin{figure}
\begin{center}
\includegraphics[width=0.95\textwidth]{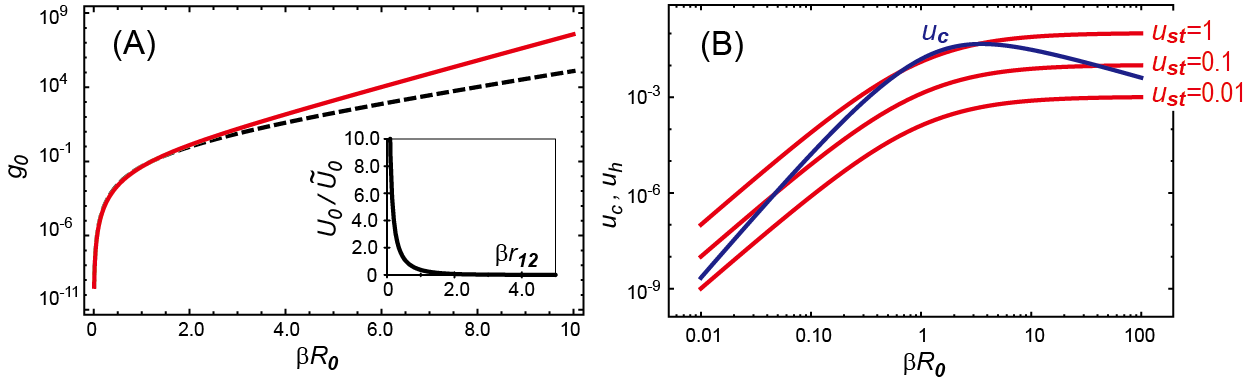}
 \caption{
 (A) \NY{Semilogarithmic plot of} the normalized concentration-mediated interaction as a function of the
 normalized size of the drop, $\beta R_0$.
 The solid (red) line shows $g_0 (\beta R_0)$\NY{, as in (\ref{uint0}),} and the
 dashed (black) line corresponds to the result under the far-field
 approximation.
 \NY{
 (inset) The interaction {\it potential} \SYYYY{$U_0/\tilde{U}_0=k_0(\beta r_{12})$} as a function of the distance
 between the two drops when $A_1 A_2>0$. 
 }
 (B) \NY{Log-log plot} of the typical hydrodynamic $u_h$ and concentration-mediated $u_c$
 interactions for \NY{drops of different normalized sizes ($\beta R_0$)}.
 The interactions are evaluated at the characteristic length scale $r_{12} = 2 R_0 + \beta^{-1}$.
 For the hydrodynamic interaction, the steady velocity \NY{of the second
 drop without interactions is decreased from
 the top to the bottom line}. 
 \label{fig.int.spherical}
 }
\end{center}
\end{figure}

The velocity in \NY{(\ref{vel.sum})} is expressed by the sum of the
velocity \NY{due to the} normal and tangential forces in (\ref{stokes.force.eq}) on the surface of the drop.
Each force is the sum of the \NY{force for an} isolated
drop and the contribution from the interaction:
\begin{align}
{\bf u}_0 + {\bf u}_c 
&=
{\bf u}_{1} + {\bf u}_{2}
\label{veocity.drop}
\end{align}
where ${\bf u}_1$ and ${\bf u}_2$ are \NY{the contributions} from the normal and
tangential forces, respectively.
\NY{The velocities can be decomposed as follows:}
\begin{align}
{\bf u}_{1} 
&=
{\bf u}_{1}^{(0)}
+ {\bf u}_{1}^{{\rm int}}
\\
{\bf u}_{2} 
&=
{\bf u}_{2}^{(0)}
+ {\bf u}_{2}^{{\rm int}}
\end{align}
\NY{
The Stokes equation
(\ref{stokes.force.eq}) is solved by using the Oseen tensor,
}
\begin{align}
\mathsf{T}_{ij}  
&=
\frac{1}{8\pi \eta}
\left[
\frac{1}{r} \delta_{ij}
+ \frac{x_i x_j}{r^3} 
\right]
\end{align}
\NY{
and the interaction can be expanded with respect to the magnitude of the
velocity of the second drop corresponding to each order in the expansion of
(\ref{cI.pair}):
}
 ${\bf u}_{1}^{{\rm int}} = {\bf u}_{1}^{{\rm int},0} + {\bf
 u}_{1}^{{\rm int},1} + {\bf u}_{1}^{{\rm int},2} + \cdots $.
The lowest-order contribution from the interaction between spherical
drops to the velocity is
\NY{
obtained from (\ref{cI0.pair}) and (\ref{Qn02b}):
}
\begin{align}
 u_{i,1}^{{\rm int},0}
&=
\frac{ \gamma_c R_0}{\Omega}
\int da \int da'
n_i (a)
{\sf T}_{jk} (a,a')
n_j (a)
n_k (a')
\left(
-\frac{2}{R_0}
\right)
 c_I^{(0)}(a')
\nonumber \\
&=
- \frac{64  \gamma_c {R_0}^2 A_2 a_{1,0}^{(1)}}{15 \eta D }
\int_{0}^{\infty}
dq q G_q
j_1(q R_0)
j_1 (q r_{12})
j_1 (q R_0)
N_i (\theta_{12},\varphi_{12})
\label{ui1002}
\end{align}
where $a_{1,0}^{(1)} = 3/(4\pi)$ and 
\begin{align}
\FSY{{\bf N}} (\theta_{12},\varphi_{12})
&=
\hat{{\bf r}}_{12}
=
\frac{{\bf r}_{12}}{|{\bf r}_{12}|}
=
\frac{{\bf r}_{2,G} - {\bf r}_{1,G}}{|{\bf r}_{2,G} - {\bf r}_{1,G}|}
\label{direction.tensor.1}
\end{align}
 is the normal vector
pointing the second drop from the first drop. 
The velocity \NY{due to the} tangential force is
 \begin{align}
  u_{i,2}^{{\rm int},0}
&=
\frac{\gamma_c R_0}{\Omega}
\int da \int da'
n_i (a)
{\sf T}_{jk} (a,a')
n_j (a)
\left[
\delta_{kl}
- n_k (a') n_l (a')
\right]
\nabla_l c_I^{(0)}
\nonumber \\
&=
\frac{ \gamma_c R_0^2}{5 \Omega \eta }
\int da' 
\left[
\delta_{ij}
- n_i (a') n_j (a')
\right]
\nabla_j c_I^{(0)}(a')
.
 \end{align}
  \NY{
 In  (\ref{cI0.pair}), the first contribution, which comes from $Q^{(0)}_1(s)$ vanishes since $(\delta_{ij}- n_i (a') n_j
 (a') ) n_j = 0$.
This is obvious since the concentration field in (\ref{Qn0.pair}) is isotropic.
 }
 The second \NY{contribution} comes from $ Q^{{\rm int}}_1(\theta,\varphi)$
 in \NY{(\ref{cI0.pair}) and} is given by
 \begin{align}
  u_{i,2}^{{\rm int},0}
&=
\frac{ 8 \gamma_c R_0^3 A_2 a_{0,1}^{(1)}}{5  \eta D}
\int_{0}^{\infty}
dq q G_q
j_1(q R_0)
j_1 (q r_{12})
j_1 (q R_0)
N_i (\theta_{12},\varphi_{12})
\label{ui2002}
 \end{align}
 where $a_{0,1}^{(1)} = (2/R_0) a_{1,0}^{(1)}$.
 \NY{
Both (\ref{ui1002}) and (\ref{ui2002}) are along the direction of the
 centreline between the two drops.
 This originates from the anisotropic concentration field created by
the {\it isotropic} field around the other drop.
Combining (\ref{ui1002}) and (\ref{ui2002}), we obtain
\begin{align}
{\bf u}_c
&=
-\nabla_{r_1} U_0 (r_{12})
\nonumber \\
 &=
-
\frac{\gamma_c A_2}{\eta D \beta^2}
k_1 (\beta r_{12})g_{0} (\beta R_0)
 \frac{{\bf r}_2 - {\bf r}_1}{|{\bf r}_2 - {\bf r}_1|}
 \label{uint0}
\end{align}
where $k_n(x)$ is the modified spherical Bessel function of the
second kind, defined as
$k_n=(-1)^n x^n (\frac{d}{x dx})^n\frac{\exp(-x)}{x}$.
The interaction may be expressed as if there is the following {\it potential}:
}
\begin{align}
 U_0 (r_{12})
&=
 \frac{16a_{1,0}^{(1)} \gamma_c R_0^2 A_2}{15 \eta D}
\int_{0}^{\infty}
dq  G_q
j_1(q R_0)
j_0 (q r_{12})
j_1 (q R_0).
\end{align}
Here we have used (\ref{besseljp1j0}).
Using (\ref{integral3sbesselj}) and (\ref{integral3sbesselj2}), we
obtain 
\begin{align}
 U_0 (r_{12})
&=
\frac{ \gamma_c A_2}{ \eta D \beta ^3 } 
g_{0} (\hat{R}_0)
 k_0 (\beta r_{12}),
  \label{uint0potential}
\end{align}
with $\hat{R}_0 = \beta R_0$.
\NY{
For given parameters, the {\it potential} decays exponentially at large
distance between the two drops as $U_0 = \tilde{U}_0 k_0(\beta r_{12})$
as shown Figure~\ref{fig.int.spherical}(A).
The magnitude of the {\it potential} $\tilde{U}_0 = \gamma_c A_2/(\eta D
\beta ^3) g_{0} (\hat{R}_0)$ depends on the parameters and the size of
the drop.
The size-dependence is explicitly given by
}
\begin{align}
 g_{0} (\hat{R}_0) 
&=
- \frac{2a_{1,0}^{(1)} \pi  }{15  \hat{R}_0^2} 
\left[ 
-2 \left( \hat{R}_0^2 + 2 \right)
 \cosh
   (2 \hat{R}_0) + 5 \hat{R}_0 \sinh (2 \hat{R}_0)+4
 \right].
 \label{g0.betaR0}
\end{align}
The plot of $g_{0}$ is shown in Figure \ref{fig.int.spherical}.
From (\ref{tauC}), the activity of the first drop is controlled by
$\gamma_c A_1$.
In order to exhibit the instability for an isolated drop, the activity
$\gamma_c A_1$ must be positive.
\NY{In that case}, the inequality $g_{0} (\hat{R}_0) \geq 0$ implies
that the {\it potential} is
repulsive when $A_1 A_2 > 0$ and attractive when $A_1 A_2 <0$.

\subsection{Far-Field Approximation}

\NY{
When the distance between two drops is significantly larger than the
size of the drops and the scaled radius is very small, we may simplify the calculation of the previous section.
For $r_{12} \gg R_0$, we have the following approximation
}
 \begin{align}
|{\bf s} - {\bf r}_{12}| 
&=
r_{12} \left[
1 + \frac{s^2}{r_{12}^2} 
- \frac{2 {\bf s} \cdot {\bf r}_{12}}{r_{12}^2}
\right]^{1/2}
\simeq
r_{12} \left[
1 
-  \frac{{\bf s} \cdot {\bf r}_{12}}{r_{12}^2}
  +
  \left(
  \frac{s^2}{2 r_{12}^2}
  - \frac{ ( {\bf s} \cdot {\bf r}_{12})^2}{2 r_{12}^4}
  \right)
  +
\cdots
  \right]
  .
 \end{align}
\NY{Instead of using (\ref{Legendre.addition}), for $\beta R_0 \ll 1$, we may
use the following expansion:} 
\begin{align}
j_0 (q |{\bf s} - {\bf r}_{12}|) 
\simeq &
j_0 (q r_{12})
 - q r_{12} j_0'(q r_{12})  \frac{{\bf s} \cdot {\bf r}_{12}}{r_{12}^2}
 \nonumber \\
 + &
 \frac{q r_{12}}{2} \left[
 j_0'(q r_{12})
 \frac{s^2}{r_{12}^2}
 + \left(
 - j_0'(q r_{12})
 + q r_{12} j_0''(q r_{12})
  \right)
 \frac{({\bf s} \cdot {\bf r}_{12})^2}{r_{12}^4}
 \right]
+ \cdots
,
\label{bessel.expand.farfield}
\end{align}
With this expansion, we will use (\ref{Qn02}) instead of (\ref{Qn02b}).
\NY{If we take only the zeroth-order term in the} expansion of the concentration field, the
velocity of the drop due to \NY{the} normal force becomes
\begin{align}
 u_{i,1}^{{\rm int},0}
&=
- \frac{16 R_0^3 \gamma_c A_2}{15 \pi \Omega \eta D}
\int da'
n_{i}^{(0)}(a') 
\int dq G_q q 
j_0(q |{\bf R}(a') - {\bf r}_{12}|)
 j_1(q R_0)
.
\end{align}
For the spherical drop, ${\bf R}(a') = R_0 {\bf n}(a')$, and therefore,
\NY{in (\ref{bessel.expand.farfield}), the terms which contain an even
number of ${\bf s}$ do not
contribute to the integral.}
The isotropic term in (\ref{bessel.expand.farfield}) does not make a
contribution for $c_I^{(0)}$.
\NY{For the lowest-order approximation}, the velocity becomes
 \begin{align}
  u_{i,1}^{{\rm int},0}
&\simeq
- \frac{16 R_0^2 \gamma_c A_2}{15 \pi \eta D}
N_i(\theta_{12},\varphi_{12})
\int dq G_q q^2 R_0 
j_1(q r_{12})
j_1(q R_0)
.
 \end{align}
 The contribution from the tangential force is
 \begin{align}
  u_{i,2}^{{\rm int},0}
&=
\frac{ \gamma_c R_0^2 A_1}{5 \Omega \eta D}
\int da' 
\left[
\delta_{ij}
- n_i^{(0)} (a') n_j^{(0)} (a')
  \right]
  \nabla_j Q_1^{{\rm int}}
\nonumber \\
&=
 \frac{4 \gamma_c R_0^3 A_2}{5 \pi \eta D}
N_{j} (\theta_{12},\varphi_{12})
\int dq G_q q^2 j_1(q R_0)
j_1(q r_{12})
.
 \end{align}
 \NY{
 Under the far-field approximation, 
we obtain the interaction ${\bf u}_c \sim  k_1(\beta
r_{12}) {\bf N}$, which is similar to that of (\ref{uint0}), and the {\it potential} $U_0 \sim
k_0(\beta r_{12})$, which is similar to that of (\ref{uint0potential}),
 although we have a different functional form for
$g_0(\beta R_0)$.
Figure~\ref{fig.int.spherical} (A) shows $g_0 (\beta R_0)$ in (\ref{uint0})
 and that obtained under the far-field approximation.
When $\beta R_0 \ll 1$, the two results agree, although they deviate
 when $\beta R_0 \gg 1$, since in that case, the far-field expansion is not justified.
}

We \NY{can analytically} confirm that
(\ref{uint0})  approaches the above expression of $  u_{i}^{{\rm
int},0}$ in the far field limit $r_{12}\gg R_0$ and \NY{$\beta R_0 \ll
1$, by using the following
relation, which holds in the far-field limit:}
\begin{eqnarray}
3\int
 dqqG_{q}\left(j_{1}\left(qR_{0}\right)\right)^{2}j_{1}\left(qr_{12}\right)
 & \simeq & R_{0}\int dq
 q^{2} G_{q}
 j_{1}\left(qR_{0}\right) 
 j_{1}\left(qr_{12}\right).
\end{eqnarray}

\NY{We can systematically compute the terms that are higher order with
 respect to the magnitude of the velocity of the second drop.
For the first-order term in the expansion}, the velocity is expressed as 
 \begin{align}
  u_{i,1}^{{\rm int},1}
&=
- \frac{8 R_0 \gamma_c A_2}{15  \Omega \eta D^2}
u_j ^{(2)}
\int da'
n_{i}^{(0)}(a') 
\pdiff{}{s_j}
Q_2^{{\rm int}}
.
 \end{align}
Similarly, the other higher order terms in \NY{the} expansion contain
\NY{higher derivatives} with respect to $s_i$. 
At the far-field limit, the first term in (\ref{bessel.expand.farfield}) does not depend on
$s_i$ and therefore\NY{, in the higher-order terms, the gradient with respect to $s_i$ vanishes}.
At the next order, $Q_n^{{\rm int}}$ is linear in ${\bf s}$ and
therefore the higher order terms do not contribute to the velocity.
The higher-order terms start to appear \NY{beginning with} the third term in (\ref{bessel.expand.farfield}).
The same argument \NY{can also be} applied to the tangential force.
\NY{From the symmetry, this} term should vanish for the second term in (\ref{bessel.expand.farfield}).
Indeed, the integral has \NY{an} odd number of normal \NY{vectors} and thus
it vanishes.
We note that \NY{although we have
considered only a spherical drop, in the general case, the shape of the first drop may affect its velocity.} 
\NY{
Consequently, the interaction cannot be  expressed in the simple form as a
potential, as in (\ref{uint0}).
}

\section{Collision of two particles}
\label{sec.amp.eq}

In the previous sections, we have discussed two-body \NY{interactions}.
The results give kinetic rules for \NY{the} position (${\bf
x}^{(\alpha)}$) and velocity (${\bf u}^{(\alpha)}$) of the $\alpha$th
drop ($\alpha = 1,2$). 
We assume there is no viscosity contrast, that is, $\eta^{(o)} = \eta^{(i)}$.
The kinetic equations  are
\NY{
\begin{align}
\odiff{{\bf x}^{(1)}}{t}
&=
 {\bf u}^{(1)}
  \label{collision.theory.x}
 \\
m \odiff{{\bf u}^{(1)}}{t}
&=
(\tau-\tau_c) {\bf u}^{(1)}
- g |{\bf u}^{(1)}|^2 {\bf u}^{(1)}
 + \tau_c \left(
 {\bf u}_c + {\bf u}_h
 \right) ,
 \label{collision.theory.u}
\end{align}
}
where the interactions due to concentration overlap and hydrodynamics
are, respectively,
\begin{align}
{\bf u}_c 
&=
- \nabla_{{\bf r}_1} U_{0} (r_{12})
 =
 - \frac{\gamma_c A_2}{\eta D \beta^2}
 g_0 (\beta R_0)
 k_1( \beta r_{12}) \FSY{\bf N}
  \label{two.drop.uc}
\end{align}
\begin{align}
{\bf u}_h 
&=
\left( \frac{R}{r_{12}} \right)^3
\left[
- \frac{1}{2} \delta_{ij}
+ \frac{3}{2} {\bf N} {\bf N}
\right] 
\cdot
{\bf u}^{(2)}
 +
 \mathcal{O} \left(
\left( \frac{R}{r_{12}}\right)^2
 {\bf S}^{(2)} \cdot {\bf N}
 \right)
 \label{two.drop.uh}
\end{align}
\NY{
where the directional vector is from the first drop and points toward the second
drop} (\ref{direction.tensor.1}).
The coefficients $m$, $\tau$, and $g$ are found in
\citep{Yabunaka:2012}.
\NY{Note that} (\ref{collision.theory.u}) is an equation for velocity
since \NY{the} {\it
effective mass} $m$ has \NY{the} dimension of time.
The equation of motion for the second drop is obtained by interchanging
the indices $ 1 \leftrightarrow 2$.
$S_{ij}$ is \NY{the} {\it dipolar} concentration distribution created around the
drop, 
$S_{ij} = (R_0 /\Omega)\int  \left[ n_i(a) n_j(a) - (1/3)\delta_{ij} \right] c(a) da$.
This second moment of the concentration field arises both from
ellipsoidal deformation and self-propulsion \citep{Yoshinaga:2014}.
Since we consider a spherical drop and \NY{assume the system is} 
close to the drift bifurcation point, $S_{ij} \sim \epsilon^2$, the
contribution of the second term in (\ref{two.drop.uh}) is negligible.

 \NY{
The sign of the coefficients is
$ - (\gamma_c A_2)/(\eta D \beta^2) g_0 (\beta R_0)
 k_1( \beta r_{12})
 <0 $ when $\gamma_c A_2 >0$.
 From (\ref{tauC}) and (\ref{g0.betaR0}), the self-propelled first drop
 (above the drift bifurcation) feels the interaction {\it potential}
 created by the second drop. 
 The two drops repel each other
 when both produce or both consume chemicals.
 When the chemical reactions of the two drops have opposite signs, the
 interaction has the opposite sign, and the two
 drops are mutually attracted.
 The interaction decays exponentially, as shown in Figure
 \ref{fig.int.spherical}(A).
 As the size of the drop increases, the concentration gradient
 associated with the concentration-mediated
 interaction becomes stronger at the fixed position.
 When the two drops approach, the interaction is best evaluated at
 $\beta^{-1}$, which is outside the interface of the drops.
 For a larger drop, the concentration field is strongly screened, and
 the interaction becomes weaker.
Therefore, the concentration-mediated interaction $u_c$ is most effective at $\beta
R_0 \simeq 1$.
 In Figure \ref{fig.int.spherical}(B), the hydrodynamic interaction ${\bf
u}_h$ evaluated at the distance of the characteristic length $r_{12} = 2 R_0
+ \beta^{-1}$ is shown for different values of the steady velocity.
This distance corresponds to the situation in which the gap between the two
drops is $\beta^{-1}$.
As the system gets closer to the critical point, the steady velocity
decreases, and the hydrodynamic interaction becomes weaker.
On the other hand, the leading order of the interaction mediated by the concentration is
independent from the steady velocity.
Therefore, near the critical point, the interaction is
dominated by the concentration field and not by the hydrodynamic interaction.
 }

 \NYY{
 For a head-on collision,
the relative position $\xi = z^{(1)} - z^{(2)}$ between the
two drops is obtained from
(\ref{collision.theory.x}) and (\ref{collision.theory.u}):
 \begin{align}
m  \ddot{\xi}
  &=
  \dot{\xi}
  \left(
  \tau-\tau_c
  - \frac{g}{4}  \dot{\xi}^2
  \right)
  - \tau_c \left(
  2 U_0'(\xi)
  +   \left( \frac{R}{\xi}\right)^3 \dot{\xi}
  \right)
  \label{interaction.reduced.relative}
 \end{align}
 where $\dot{\xi} = d\xi/dt$.
 The hydrodynamic interaction is
 repulsive before the collision.
 However, after the collision, and when} the
 two drops move away from each other, the interaction becomes
attractive.
This \NY{contrasts with the behaviour seen in an} isotropic concentration-mediated interaction.
\NY{Near the critical point,} the steady velocity of a drop is small,
 \NY{and thus} the interaction created by the concentration field is stronger
 than that \NY{created} by the hydrodynamics.
  \NYY{
 Trajectories and velocity of the solution of
 (\ref{interaction.reduced.relative}) are shown in
 section~\ref{sec.numerical.interaction} together with the numerical
 results.
 }
 \NY{
Although the assumptions that we have made in the calculation of the
interactions are not completely justified especially during collision,
we will show in the following sections that  this is in semi-quantitative agreement with
our numerical results.
}

\begin{figure}
\begin{center}
\includegraphics[width=0.60\textwidth]{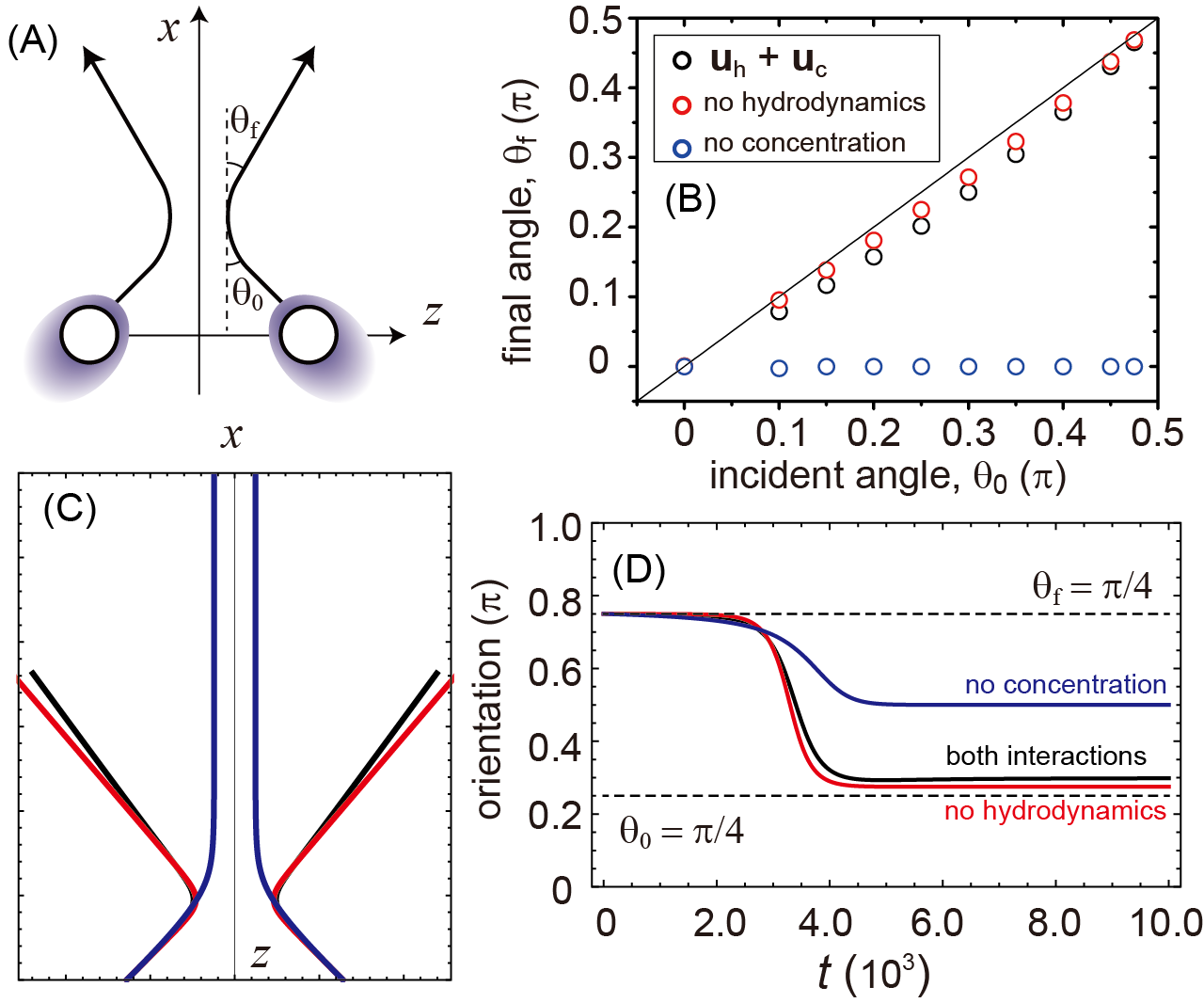}
 \caption{
 \NYY{
 (A) A schematic representation of a collision, (B) the incident,
 $\theta_0$, and
 final, $\theta_f$, angles for the solution of (\ref{interaction.reduced.relative2a})
 and (\ref{interaction.reduced.relative2b}) (black), without the
 hydrodynamic interaction (red), and without the concentration-mediated
 interaction (blue),  (C) the trajectories during the collisions for
 $\theta_0= \pi/4$, and
 (D) the direction of motion during the collisions for
 $\theta_0= \pi/4$.
 }
\label{fig.collision.angle}
}
\end{center}
\end{figure}

\NYY{
When the motion of two drops is confined in the $xz$-plane, and the
collision has a symmetry with respect to the $x$-axis,
the dynamics are expressed by $\xi = z^{(1)} - z^{(2)}$ and $\rho =
x^{(1)} + x^{(2)}$, as follows:
 \begin{align}
m   \ddot{\xi}
  &=
  \dot{\xi}
  \left(
  \tau-\tau_c
  - \frac{g}{4}
  \left(
   \dot{\xi}^2
  +
     \dot{\rho}^2
  \right)  
  \right)
  - \tau_c \left(
  2 U_0'(\xi)
  +   \left( \frac{R}{\xi}\right)^3 \dot{\xi}
  \right)
    \label{interaction.reduced.relative2a}
  \\
  m   \ddot{\rho}
  &=
  \dot{\rho}
  \left(
  \tau-\tau_c
  - \frac{g}{4}
  \left(
   \dot{\xi}^2
  +
     \dot{\rho} ^2
  \right) 
  \right)
  - \frac{\tau_c}{2} 
  \left( \frac{R}{\xi}\right)^3 \dot{\rho}
  .
  \label{interaction.reduced.relative2b}
 \end{align}
 Two drops collide with an incident angle $\theta_0$ and a final angle
 $\theta_f$ (Figure~\ref{fig.collision.angle}(A)).
 Trajectories of the solution of (\ref{interaction.reduced.relative2a})
 and (\ref{interaction.reduced.relative2b}) for $\theta_0=\pi/4$ are shown in
 Figure~\ref{fig.collision.angle}(C).
 The parameters are chosen to be the same as the numerical simulations
 for $\eta=2.3$. 
 During the collision, the direction of motion changes from the incident
 angle, $\theta_0$, and it reaches the final angle, $\theta_f$
 (Figure~\ref{fig.collision.angle}(D)).
 The final angle is dependent on the incident angle, as shown in
 Figure~\ref{fig.collision.angle}(B).
 When the incident angle is between $0$ and $\pi$, the final angle is
 smaller than the incident angle.
The hydrodynamic interaction enhances this effect; the final angle is
 less smaller than the incident angle if we eliminate the hydrodynamic interaction.
On the other hand, if we eliminate the concentration-mediated interaction, two
drops always align.
}

\section{Numerical simulations: single drop}
\label{sec.numerical.isolated}

We numerically solve \NY{(\ref{DphiDt}), which gives the dynamics of a
drop and (\ref{stokes.force.eq}), which gives the flow
field.}
\NY{We used the following equation for the dynamics of the
concentration field similar to (\ref{con.N}):}
\begin{align}
\pdiff{c}{t} 
+ {\bf v} \cdot \nabla c
&=
D\nabla^2 c - \kappa c
+ 
\frac{1}{2} A  \left(
\phi ({\bf r})+1
 \right)
 .
\label{con.numerics}
\end{align}
We \NY{assumed an} axisymmetric system \NY{in which} the motion of the drop \NY{was} confined
along the $z$-axis.
The \NY{entire} space \NY{was} discretized \NY{in cylindrical coordinates,} with $N_z
= 96$ and $N_r = 48$ mesh points in \NY{the} $z$- and radial ($r$-)
directions\NY{, respectively}.
The mesh size \NY{was} chosen \NY{to be} $\Delta z = \Delta r=1$\NY{,} and the time step \NY{was}
$\Delta t = 0.002$.
\NY{Equations} (\ref{con.numerics}) and (\ref{DphiDt}) were discretized
\NY{by} using the
forward Euler method.
We imposed \NY{a} periodic boundary condition in \NY{the} $z$-direction,
and, at $r=N_r$, \NY{we imposed} the slip boundary condition $\partial_r v_z=0$ and $v_r=0$, the non-wetting boundary condition  $\partial_r \phi =0$\NY{,} the no-flux boundary condition $\partial_r \frac{\delta f}{\delta \phi}=0$\NY{,} and $\partial_r  c =0$.
The Stokes equation \NY{was} solved \NY{by using} the relaxation method;
\NY{for a} given force
${\bf f}({\bf r})$, (\ref{stokes.force.eq}) \NY{was} solved by introducing the
virtual time derivative $d
{\bf v}/dt$ and relaxing until \NY{a steady state was obtained}.
At each step in the virtual time domain, the pressure \NY{was} relaxed to
the steady value so that the incompressibility condition \NY{was} satisfied.
\NY{For the} discretization, we employed the staggered lattice
method\NY{, in which} the $r$ and $z$ \NY{components of the flux were} defined at the lattice \NY{points} $((i +1/2) \Delta r ,j \Delta z )$ and $(i  \Delta r ,(j+1/2) \Delta z )$, respectively, for $i=0,...,N_{r}-1$ and $j=0,...,N_{z}-1$.
 In order to quickly prepare initial conditions that are stationary under (\ref{con.numerics}) and (\ref{DphiDt}) with $\bf{v}=0$, we solved the following equations
\begin{align}
\pdiff{c}{t'} 
&=
\alpha \left[ D\nabla^2 c -\kappa c
+ 
\frac{1}{2} A  \left(
\phi ({\bf r})+1
\right) \right],
\label{con.numerics-2}
\end{align}
\begin{align}
\pdiff{\phi}{t'} 
&=
-\frac{\delta F}{\delta \phi}+\left< \frac{\delta F}{\delta \phi} \right>,
\label{DphiDt2}
\end{align}
with $\alpha=20$ and $\Delta t'=0.01$ until $t'=160$, starting from $c=0$ and $\phi=-\tanh (R - |{\bf r} - {\bf r}_{G,1}|)+0.05$ at $t'=0$. 
We \NY{added} a \NY{small-amplitude} noise to this initial condition
\NY{in order} to investigate \NY{the} self-propelled motion of a drop with spontaneous symmetry breaking.

First, we \NY{dropped} the advection term in  (\ref{con.numerics})
\NY{in order} to \NY{directly compare our} numerical results with the
theoretical \NY{predictions of} \citep{Yabunaka:2012}.
We \NY{varied} the viscosity $\eta$ to realize self-propulsive motion under
fixed $R=16$, $B_0=0.2$, $B_1=0.5$,  $D=0.5$, $A=0.08$, $\kappa=0.005$,
and $L=1$.
Here $\beta= \sqrt{\frac{\kappa}{D}}=0.1$.
 We confirmed that this choice of parameters satisfies the assumptions
 \NY{listed following} (\ref{amp.eq.single}).
\NYY{
 With these parameters, the critical value of the
 viscosity is \SSY{theoretically predicted as} $\eta_c \simeq 1.742$ \NY{when} using (\ref{tauC}).
\SSY{We will discuss possible reasons for the discrepancies between the numerical results and the theoretical predictions in detail in section \ref{sec.numerical.interaction}. }
 }
In order to estimate the critical value, we need to evaluate the surface
tension and $\gamma_c$.
When the interface is sharp and the value of $c({\bf r})$ at the
interface is unique, then  $\gamma= \frac{2\sqrt{2B(c)}}{3}$ and $\gamma_c = \frac{\sqrt{2}B_1}{3\sqrt{B(c)}}$.
However, since we use a diffuse interface model, the concentration at
the interface {\it region} varies in space.
The surface tension is
\begin{align}
\gamma 
&=
\int B(c)
\left(
\pdiff{\phi}{n}
\right)^2
dn
\label{gamma.numerical}
\end{align}
where $\partial/\partial n $ is the spatial derivative along the
\NY{direction} normal to the interface.
We numerically \NY{estimated} (\ref{gamma.numerical}) and \NY{compared
the results} with
(\ref{gamma}) to obtain $\gamma_c\sim 0.510$.

\begin{figure}
\begin{center}
\includegraphics[width=0.40\textwidth]{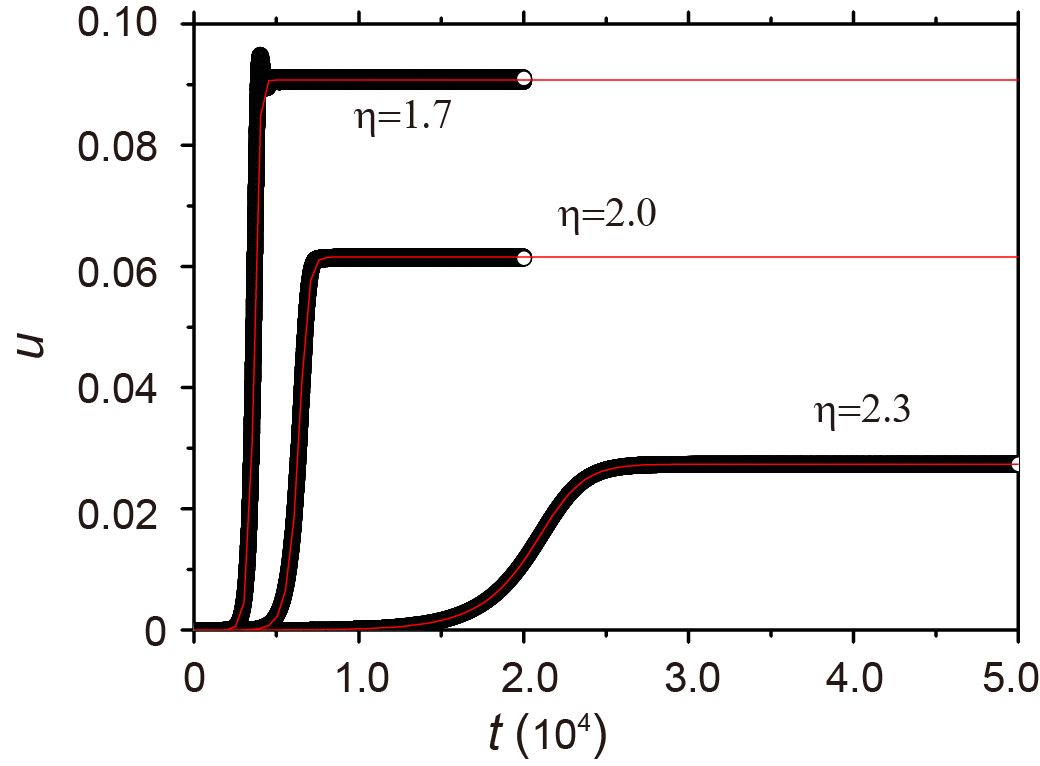}
\caption{
Self-propulsive velocity of a single drop. 
 The velocity of the drop as a function of time is shown \NY{for
 numerical simulation (circles) and theory in  (\ref{steady.vel.single})
 (line).}
\label{fig.numerics.isolated.V}
}
\end{center}
\end{figure}

\begin{figure}
\begin{center}
\includegraphics[width=0.70\textwidth]{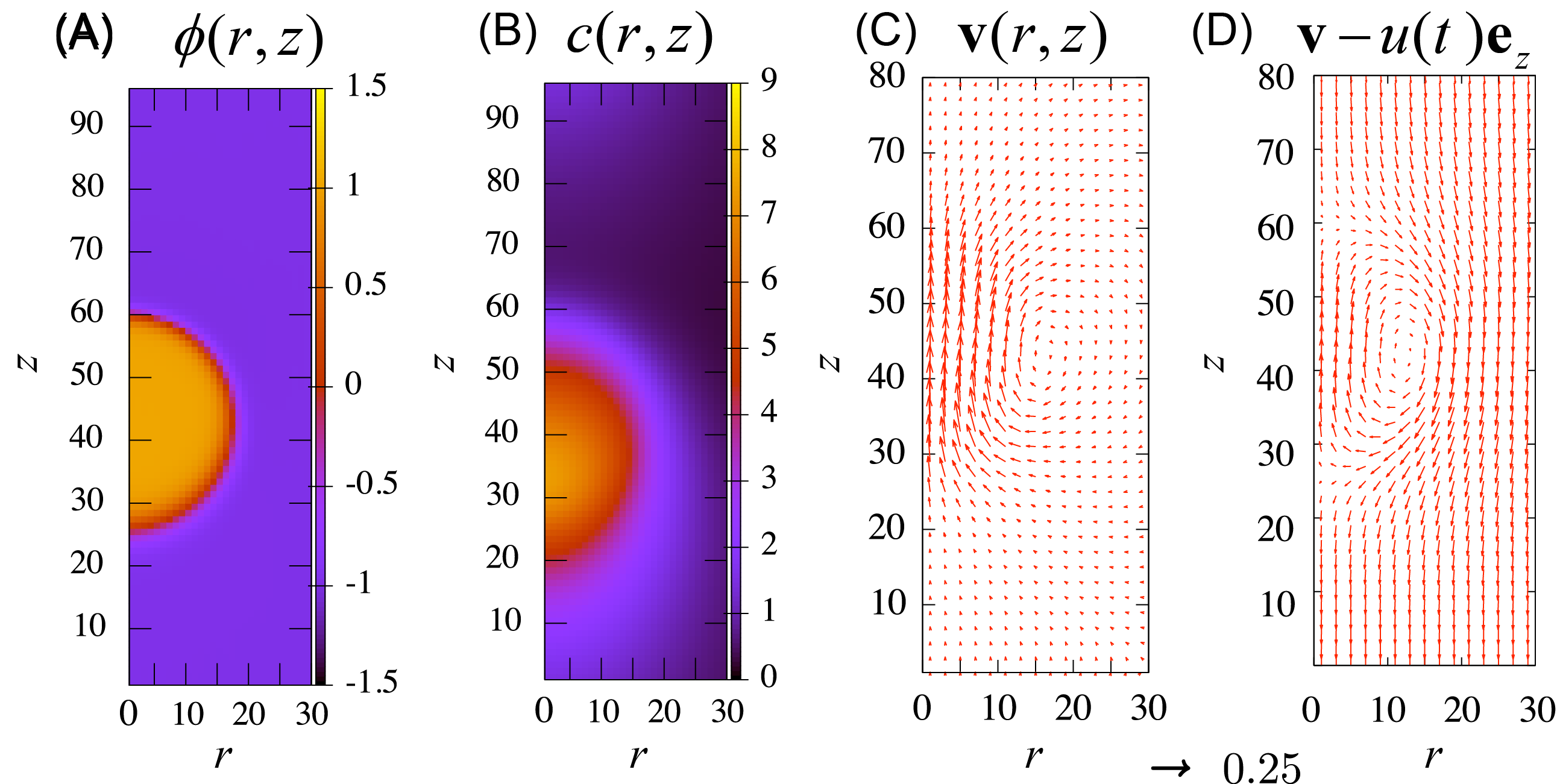}
 \caption{
 \NY{For a drop at steady state with $\eta=1.5$ and $t=3168$: (A)
 $\phi({\bf r})$, (B) the concentration field $c ({\bf r})$, and (C, D) the velocity field ${\bf v} ({\bf
 r})$ in the laboratory frame (C), and in the drop frame (D).}
 The self-propulsive velocity is $u=0.109$. 
\label{fig.numerics.isolated}
}
\end{center}
\end{figure}

When $\eta < \eta_c$, the stationary state becomes unstable\NY{,} and the drop
starts to move.
The velocity of the drop gradually increases \NY{until it reaches a steady
state, as shown} in Figure \ref{fig.numerics.isolated.V}.
During the self-propulsive motion, the concentration field is distorted
around the drop\NY{, as} shown in Figure \ref{fig.numerics.isolated}.
Clearly, the two \NY{centres of mass (that of the drop and that of the concentration) are
shifted,} and thus the symmetry \NY{is broken for} the $\pm z$ directions.
The velocity field during the motion is shown in Figure
\ref{fig.numerics.isolated}.
The velocity field in the drop frame shows \NY{a} circular flow \NY{that corresponds}
to the $l=1$ mode in Figure \ref{fig.flow.schematic}.
\NY{Around the drop, the} velocity field decays faster than $1/r$\NY{,} suggesting that the
motion is generated by the source dipole and not by \NY{a} Stokeslet.

\begin{figure}
\begin{center}
\includegraphics[width=0.95\textwidth]{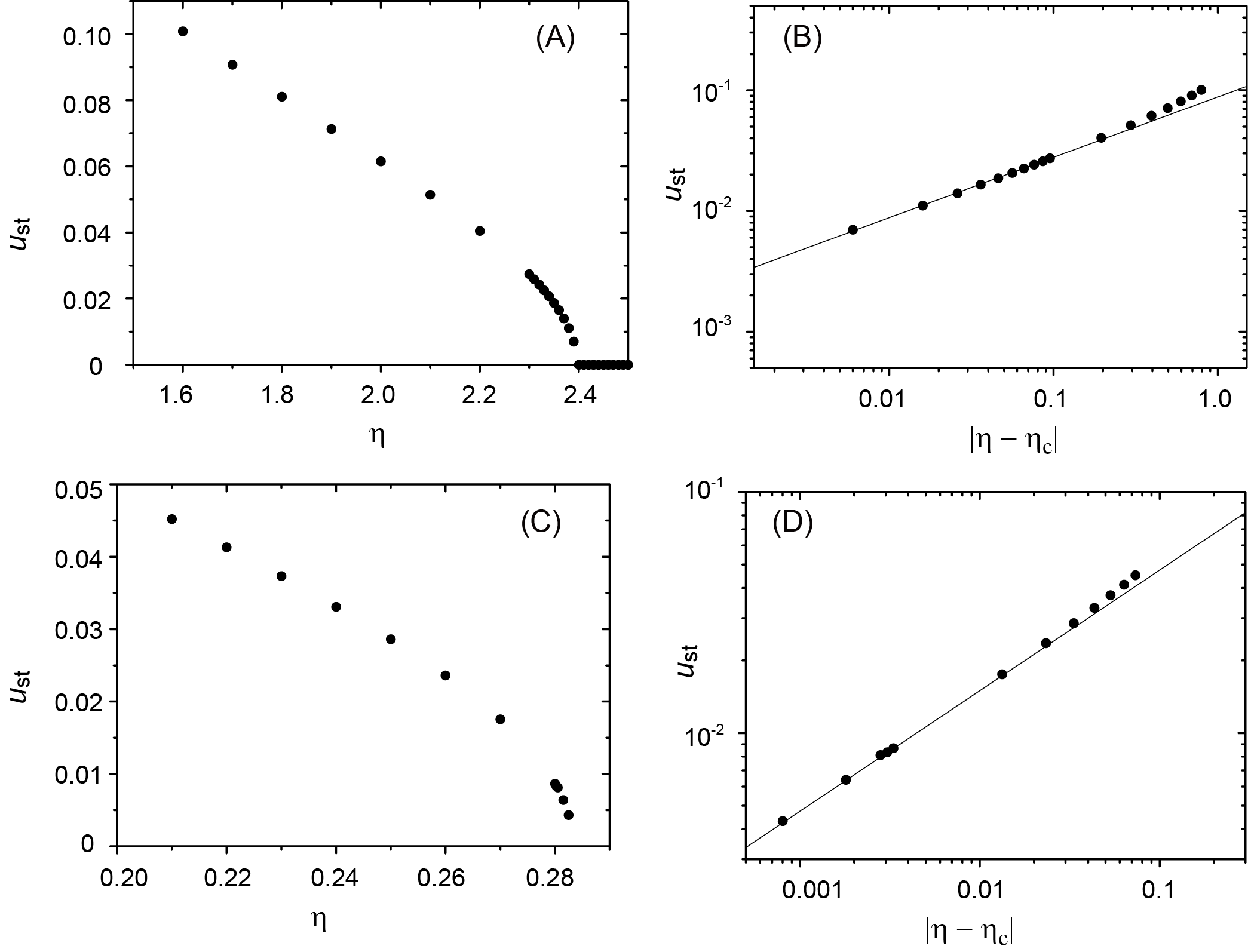}
\caption{
 The steady velocity $u_{\rm st}$ with varying viscosity $\eta$ ($\tau_c$).
 (A,B) \NY{Without advection of the third dilute component,} and (C,D) with advection.
The log-log plots of the steady velocity and the distance from the
 critical points are shown in (B) and (D).
 The solid lines show the exponent $u_{\rm st} \sim |\eta - \eta_c|^{1/2}$.
 \label{fig.numerics.isolated.bifurcation}
}
\end{center}
\end{figure}

\begin{figure}
\begin{center}
\includegraphics[width=0.95\textwidth]{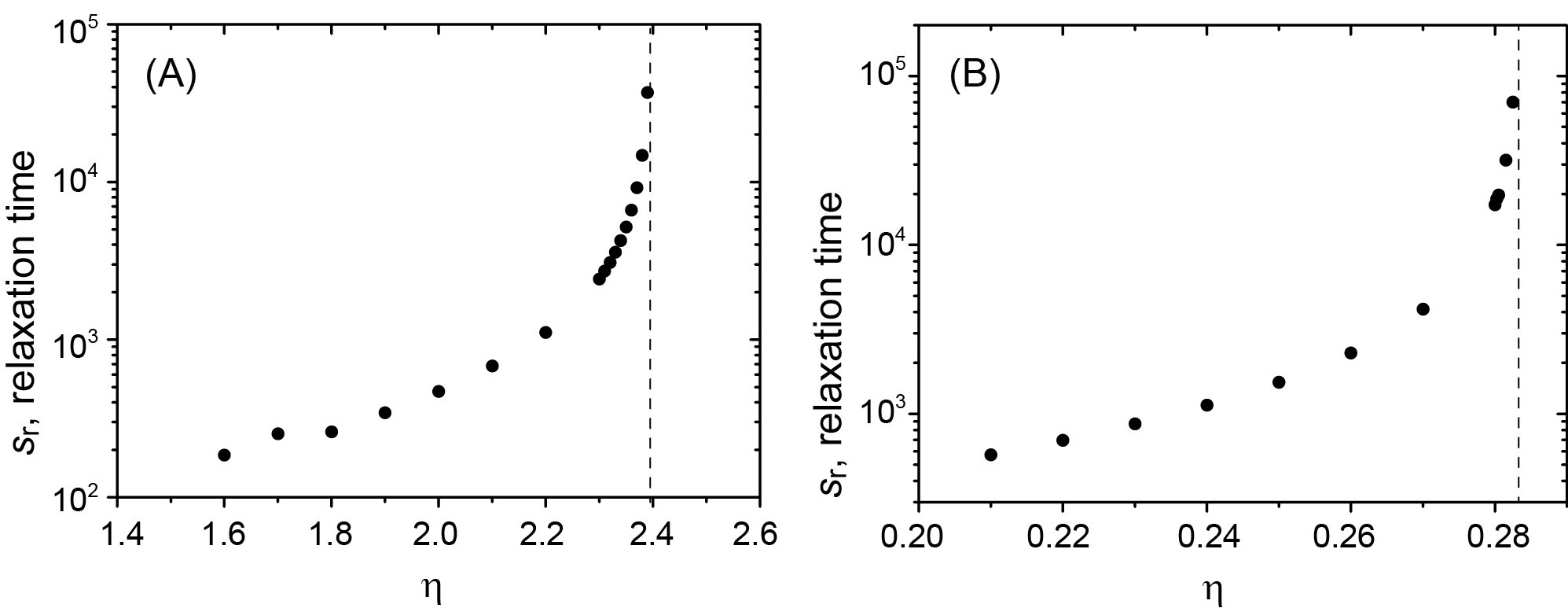}
\caption{
 The relaxation time to reach the steady velocity.
  (A) \NY{Without advection of the third dilute
 component}, and (B) with advection.
 The vertical dashed lines correspond to the critical viscosity.
 \label{fig.numerics.isolated.relaxation.time}
}
\end{center}
\end{figure}

\NY{When the viscosity $\eta$ is
close to the critical value,} the relaxation of the velocity is monotonic
and well fitted with (\ref{steady.vel.single2}).
This is consistent with the theoretical result of
(\ref{steady.vel.single}).
\NY{When} $\eta$ \NY{is} much \NY{smaller than} the critical value, however, the
relaxation of the velocity is not monotonic but \NY{has} a small
oscillation\NY{, as seen}
for $\eta=1.7$ in Figure \ref{fig.numerics.isolated.V}.
This may arise from \NY{the involvement of} an additional time
scale\NY{, and the truncation of} the expansion
of (\ref{con.sol.fourier.pari}) is not completely satisfied. 
\NY{
The steady velocity and the relaxation time are plotted in Figures
\ref{fig.numerics.isolated.bifurcation}.
As the viscosity $\eta$ decreases, that is, as $\tau_c$ decreases, the
self-propulsive speed increases.
As suggested in Figures~\ref{fig.numerics.isolated.bifurcation}(B) and
(D), close to the critical point, the speed increases as $u \sim |\eta_c -
\eta|^{1/2}$, which is predicted by the theory in (\ref{steady.vel.single}).
Figure \ref{fig.numerics.isolated.relaxation.time} shows the relaxation
time of the numerical simulations.
As the viscosity approaches the critical value from below, the relaxation time
diverges.
This behaviour is also consistent with our theory in (\ref{steady.vel.single2}).
}

\begin{figure}
\begin{center}
\includegraphics[width=0.95\textwidth]{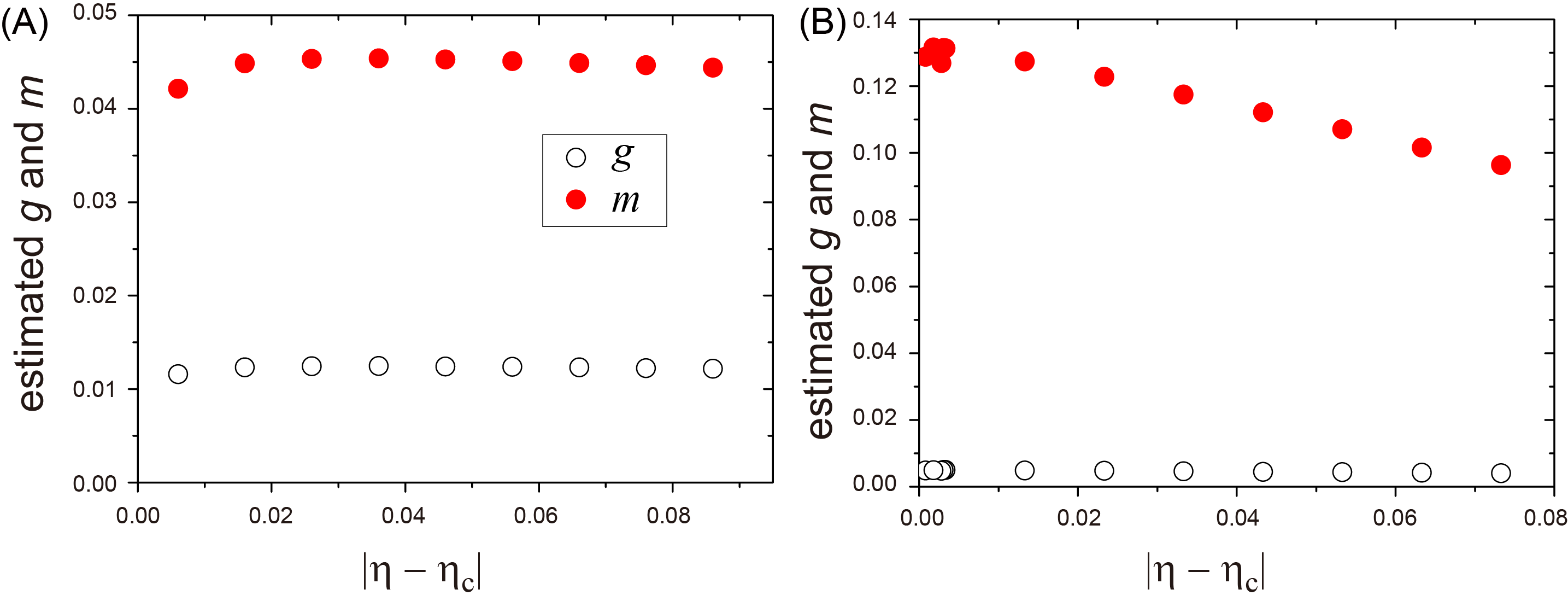}
\caption{
 Numerically estimated coefficients $g$ and $m$.
  (A) \NY{Without advection of the third dilute component,} and (B) with advection.
 \label{fig.numerics.isolated.coefficients}
}
\end{center}
\end{figure}

We evaluate the scaled coefficients \NY{$m$, $\tau$, and $g$}
($\hat{m},\hat{\tau},\hat{g}$ in \citep{Yabunaka:2012})  \NY{by
comparison with the numerical results, as follows.}
In terms of the non-scaled coefficients \NY{$M$, $T$, and $G$} ($m,\tau,g$ in
\citep{Yabunaka:2012}), the relaxation time and the steady velocity
are expressed, respectively, as
\begin{equation}
 s_r = \frac{M}{T-1}
\end{equation}
\begin{equation}
 u_{st}=\sqrt{\frac{T-1}{G}}
  .
\end{equation}
Since $T\sim\eta^{-1}$, the above expression \NY{for} $s_r$ predicts the
divergence of the relaxation time near the threshold, which \NY{is in agreement} with
the numerical results \NY{in Figure \ref{fig.numerics.isolated.relaxation.time}}.
Using $\eta_{c}=2.395$, we estimate
$T-1\sim0.19$ for $\eta=2.0$.
From the above equations, we estimate $G\sim49.4$ and $M\sim96.9$.
\NY{After} rescaling the parameters \NY{of} \citep{Yabunaka:2012}, we obtain
\begin{equation}
 g=\tau_{c}\left(D\beta\right)^{2}G=0.0112,
\end{equation}
\begin{equation}
 m=MD\beta^{2}\tau_{c}=0.044.
\end{equation}
These values of $m$ and $g$ agree with \NY{the} theoretical
predictions in  \citep{Yabunaka:2012} with $\beta R=1.6$.
In the same way, we evaluated $g$ and $m$ for other values of $\eta$
near $\eta_c$\NY{, as shown in Figure~\ref{fig.numerics.isolated.coefficients}.
We found that they are almost constant, which is consistent with \SSY{the}
theoretical predictions.
}

The advection term in (\ref{con.numerics}) does not \NY{change} the
qualitative \NY{features} of the transition; we found that, when $\eta < \eta_c=0.2833$, the stationary state becomes unstable\NY{,} and the drop
starts to move.
In addition, Figures \ref{fig.numerics.isolated.bifurcation} and
\ref{fig.numerics.isolated.relaxation.time} show \NY{that} the advection does not
modify the scaling \NY{behaviour} of the steady velocity near the critical
point.
Both with and without the advection term, the steady velocity grows as $|\eta
- \eta_c|^{1/2}$.
The relaxation time also \NY{diverges} near the bifurcation point.
We note that the critical point \NY{when there is advection
($\eta_c=0.2833$)} is much smaller than \NY{when there is no advection ($\eta_c=\FSY{2.395}$)}.
In \citep{Yabunaka:2012}, the effect of the advection is only treated in
the limit $\beta R \rightarrow 0$ and it is predicted that, in \NY{the} presence
of advection, the drift instability is suppressed\NY{,} but the
bifurcation \NY{behaviour will not be essentially} changed.
 Thus\NY{,} our numerical results with $\beta R=1.6$ agree
 \NY{qualitatively} with the theoretical prediction \NY{as} $\beta R
 \rightarrow 0$, \NY{even though} the theory does \NY{not directly} apply to our case.

\section{Numerical simulations: interaction}
\label{sec.numerical.interaction}

With the same numerical method \NY{that we used} in the previous section
for an isolated drop, we carried out \NY{numerical simulations for} two drops.
The \NY{entire} space \NY{was} discretized in cylindrical coordinates with $N_z
= 200$ and $N_r = 48$ mesh points in \NY{the} $z$- and radial ($r$-)
directions.
The mesh size \NY{was} chosen \NY{to be} $\Delta z = \Delta r=1$\NY{,} and the time step \NY{was}
$\Delta t = 0.002$ for $\eta>1.9$ and $\Delta t = 0.001$ for $\eta<1.9$.
We \NY{prepared} the initial condition \NY{in the same way,} but with 
$\phi=\tanh (R - |{\bf r} - {\bf r}_{G,1}|)+\tanh (R - |{\bf r} - {\bf r}_{G,2}|)-2+0.05$ with  ${\bf r}_{G,1}=29$ and ${\bf r}_{G,2}=171$.

\begin{figure}
\begin{center}
\includegraphics[width=0.70\textwidth]{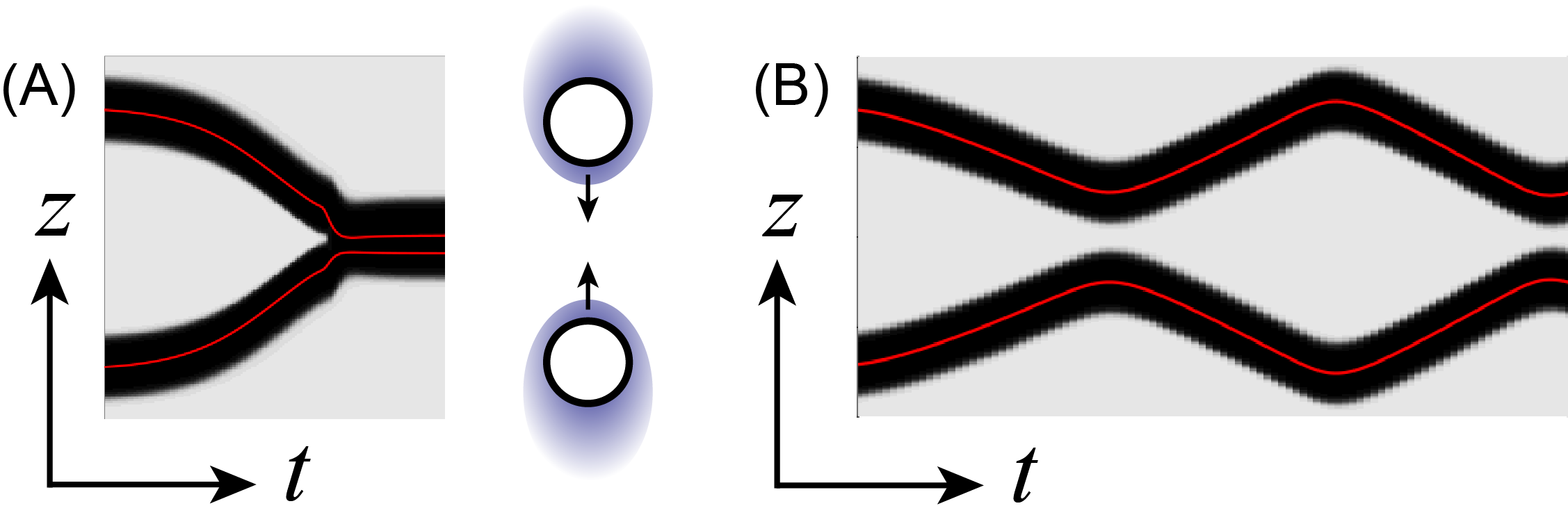}
\caption{
The trajectories of two colliding drops for (A) $\eta=1.5$ and (B)
 $\eta=2.3$.
 The slices of the field $\phi$ at $r=0$ are superposed for the
 direction of time.
 The darker (brighter) region corresponds to $\phi=1$ ($\phi=-1$).
 The solid lines show the trajectories of the \NY{centres} of the drops.
 Because of the periodic boundary condition, the drops reaching the top or
 bottom boundaries are reflected by the interactions with image drops
 outside the simulation box. 
 \label{fig.interaction.trajectoty}
}
\end{center}
\end{figure}

Figure \ref{fig.interaction.trajectoty} shows the trajectory of the two
\NY{centres} of mass of the interacting particles.
There are two distinct dynamics for \NY{the} collision: \NY{fusion, as
shown in Figure~\ref{fig.interaction.trajectoty}(A), and reflection, as
shown in Figure~\ref{fig.interaction.trajectoty}(B).}
When the viscosity $\eta$ is far below the critical point, the two
drops approach and eventually merge.
On the other hand, when $\eta$ is close to the critical point, the two
drops do not merge even when they are approaching, but they reflect and
move \NY{in} opposite directions.
Surprisingly, the latter collision is elastic\NY{,} despite the fact that the
system is dissipative.
The drops \NY{move at}
the same speed after \NY{the} collision as \NY{they did} before \NY{the}
collision.
This \NY{can} be understood from our reduced description
(\ref{amp.eq.single}), \NY{in which
{\it friction} vanishes as $\tau_c$ approaches $\tau$; note that $\tau_c$ is proportional to $\eta$ as shown in (\ref{tauC}).}
Thus, the drop behaves as if it were in a conserved system because of
the balance between dissipation and \NY{the} energy injection
\NY{associated with} chemical
production. 
This is in contrast with \NY{both} the squirmer and \NY{the} Janus
particle\NY{, since in those models, there is no {\it inertia} term} in the equation of motion. 
We note that similar elastic \NY{behaviour has been} reported for pulse
\NY{collisions in reaction-diffusion} systems \citep{Ohta:1997a,ei:2006}.

\begin{figure}
\begin{center}
\includegraphics[width=0.99\textwidth]{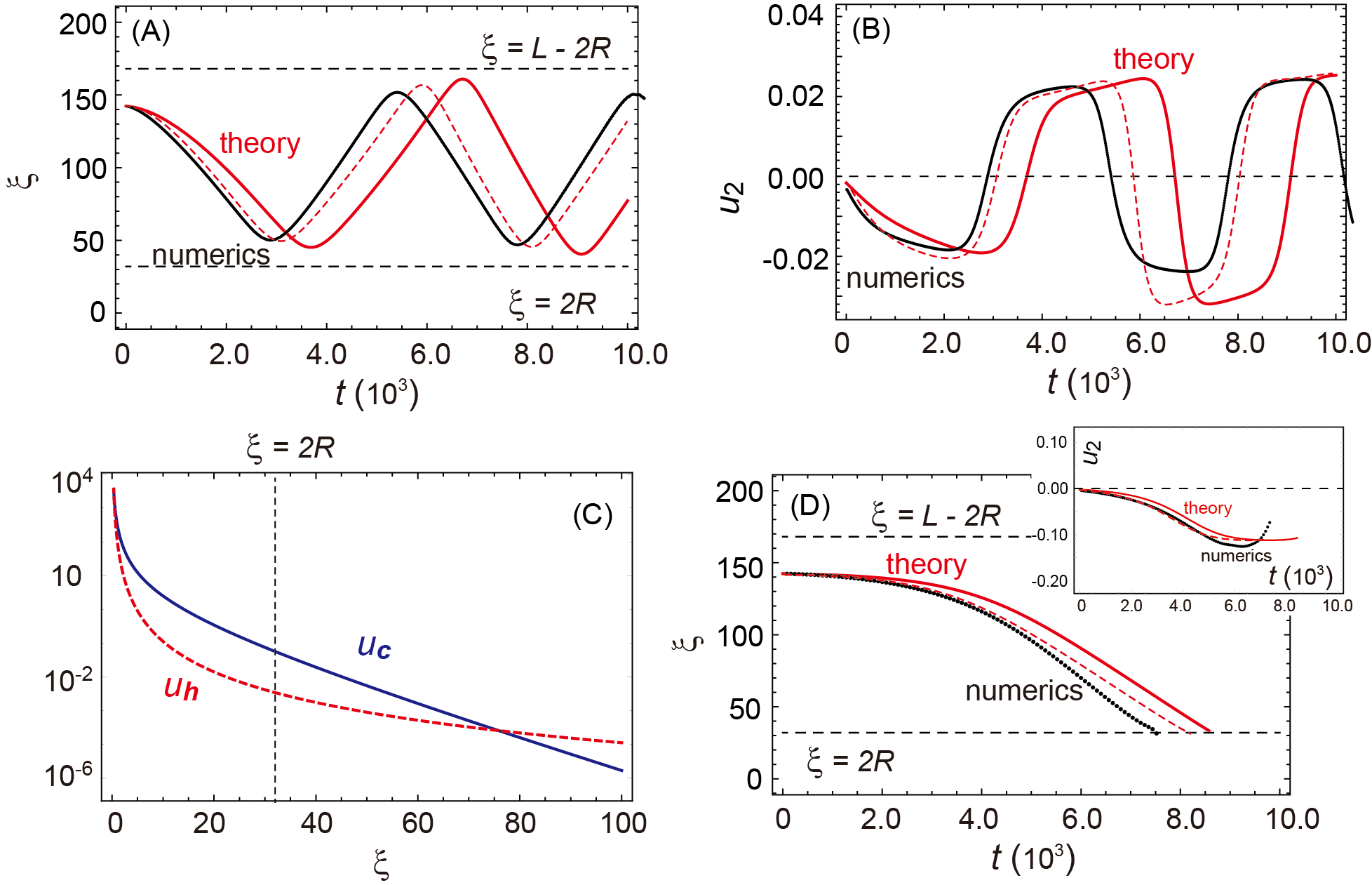}
\caption{ 
Comparison between \NY{the} reduced equations, (\ref{collision.theory.u}) and (\ref{interaction.reduced.relative}), and
 the full model for \NY{(A-C) $\eta=2.3$ and (D) $\eta=1.5$.}
 (A) The distance $\xi= |z^{(1)}-z^{(2)}|$ between two drops as a function of time, (B) velocity
 of the second drop, and (C) the dependence of \NY{the}separation distance on the hydrodynamic $u_h$ and
 concentration-mediated $u_c$ interactions\NY{, where} $u_h$ is estimated from the steady velocity $u_{\rm st}$.
 (D) The distance between two drops for $\eta=1.5$ and (inset) velocity of the
 second drop.
 \NYY{
 The dashed lines in (A), (B), and (D) are obtained from our theory
 using $R=18.5$ instead of $R=16$.
 }
 \label{fig.interaction.comparison}
 }
\end{center}
\end{figure}

\NY{
In order to clarify the origin of the interaction, we solve the reduced
equation (\ref{interaction.reduced.relative}), and compare with the full numerical
 simulations using the same initial conditions.
 }
 \NYY{
 The additional terms, which describe the interaction with image drops, are added to
 (\ref{interaction.reduced.relative}) in order to take into account the
 periodic boundary condition used in the simulation.}
 The parameters in (\ref{interaction.reduced.relative}) are obtained
 from the result of a single drop, and thus there is no fitting
 parameter in the equation.
 The result for $\eta=2.3$
 (Figure~\ref{fig.interaction.trajectoty}(B)) is shown in
 Figures~\ref{fig.interaction.comparison} (A) and (B).
 \NY{There is good agreement between the results of the reduced
 equation and those of the original model.
We found that the overall behaviour} of the evolution of $\bf{u}$ is dominated by the
concentration-mediated interaction\NY{,} although the hydrodynamic interaction
gives some correction on it.
\NY{The contributions} from the hydrodynamic ($\sim \xi^{-3}$) and
 concentration-mediated interactions ($\sim U_0'(\xi)$) \NY{are shown} in
 Figure~\ref{fig.interaction.comparison}.
 \NY{Over} most of the region, $u_c$ dominates $u_h$.
 However, the hydrodynamic interaction dominates \NY{when} $\xi \gg 80$, since the hydrodynamic (concentration-mediated) interaction decays algebraically (exponentially).
 For $\eta=1.5$ (Figure~\ref{fig.interaction.trajectoty}(A)), we
 solved (\ref{interaction.reduced.relative}), as shown in
 Figure~\ref{fig.interaction.comparison}(D) and \NY{we} found that $\xi$
 becomes \NY{smaller} than $2R$, which suggests fusion of the drops and
 agrees with the results of \NY{the} numerical simulation \NY{with} the original model.

This result is consistent with the following rough estimate of the relative
magnitudes of these two interactions:
The magnitude of the hydrodynamic interaction is estimated \NY{to be}
\begin{equation}
\left|\bf{u}_{h}\right| =
 \left(\frac{R}{\xi}\right)^{3}u_{j}^{\left(2\right)}
  \sim u_{\rm st} \left(\frac{R}{\xi}\right)^{3},
\end{equation}
where the typical self-propelling 
velocity \SYYYY{in the steady state} is given by $u_{\rm st} \sim0.02$.
The magnitude of the concentration-mediated interaction is
\begin{align}
 \left|\bf{u}_{c}\right|
 &\sim  \frac{15 D \beta}{2 \tau_c} g_0(\hat{R}_0) k_1 (\beta \xi),
\end{align}
where $g_{0}\left(\hat{R}_{0}=1.6\right)\sim0.39$.
If we set $\xi=2R$ \NY{and} $k_1 (\beta \xi) \sim 0.0167$, then ${\bf u}_{c}
\sim 0.213$ and ${\bf u}_{h} \sim 0.0025$.
This also confirms that the magnitude of the concentration-overlap-mediated interaction is larger than that of the hydrodynamic interaction.

\NY{
The discrepancy between theory and the numerical simulations arises for several
reasons.
First, our reduced description is valid only near the critical
point, which is $\eta \simeq  2.395$. We choose the parameter as close to the
critical point as possible. Nevertheless, the gap ($|\eta - \eta_c| \simeq 0.1$) would
lead to higher-order terms in (\ref{collision.theory.u}) and accordingly in
(\ref{interaction.reduced.relative}).
Second,
there is a small discrepancy between the steady-state velocity of a single drop
and that of two drops because of the difference in the system size.
We used the
parameters associated with the steady velocity and relaxation time from
the motion of a single drop.
Third, we assumed that the distance
between the two drops is large, and thus the interaction when the two
drops approach is not accurately described by the reduced
equations.
\NYY{
In addition to these reasons, a drop in the diffused-interface model has
a finite width of an interface.
In order to use a thin interface, we need to make fine discretization in
space, and
this requires a huge computational cost.
The size of a drop in our model is, therefore, not accurately given.
If we use a slightly larger size in our theory, the agreement between
the results of the reduced  equation and those of the original model
becomes better (Figure~\ref{fig.interaction.comparison}).
 From this observation, we speculate the main error arises from the lack
 of the
 accuracy of estimation of the size.
}
It is also noted that both relative position and relative
velocity are not instantaneous quantities, but history-dependent, as in
(\ref{interaction.reduced.relative}).
This is because of the effective inertia term of the reduced
equations.
Therefore, all of these errors increase with time. 
}

\section{Discussion and Summary}
\label{sec.summary}

We have developed the theory of \NY{a} collision between two self-propelled
drops driven by chemical reactions.
Close to the bifurcation point between stationary and self-propelled
states, the collision is elastic\NY{,} while away from the point, fusion
occurs.
The interactions originate from \NY{the} hydrodynamics and \NY{the} overlap of the
concentration field.
\NY{Both} interactions are repulsive during \NY{a} head-on collision if the
chemical reactions of the two drops have \NY{the} same sign \NY{(both
producing or both consuming)}.
 We found that the concentration-mediated interaction dominates the
 collision dynamics.
 Our analytical calculation is confirmed semi-quantitatively by the
 numerical results.

We stress \NY{that} inertia-like and nonlinear terms naturally appear in the
reduced description (\ref{amp.eq.single}).
 These effects are confirmed by the numerical results\NY{; the
 self-propulsion occurs above the bifurcation point at which the
 relaxation time diverges.}
 The steady velocity obtained \NY{as a function of the} distance from the critical point
 also fits with our theory.
 During \NY{a} collision of two drops, we obtain elastic \NY{behaviour} near the
 critical point.
 This is consistent with the existence of \NY{an} inertia-like term.
 The current model has no intrinsic polarity (direction)\NY{,} and therefore\NY{,}
 there is \NY{a} marked difference \NY{between its} collision dynamics \NY{and} that of the linear
 squirmer model.
 In the latter, \NY{a} change \NY{in} direction is inevitably followed by
\NY{a} rotation\NY{,} while \NY{in} the current model\NY{, a change in
the direction is instantaneous}.
 It has been argued that the competition between self-propulsion and
 \NY{the} rotational diffusion time plays a relevant role in the collective
 \NY{behaviour of} the squirmer and Janus \NY{particles} \citep{Cates:2014,Matas-Navarro:2014}.
 Our study reveals that \NY{a symmetry-breaking} swimmer may have another
 mechanism of competition, possibly between self-propulsion and \NY{the}
 effect \NY{of inertia}.
 This may lead to another phase in the collective \NY{behaviour} of
 self-propelled particles.

 \NY{
Although we have focused on two-body interactions, the behaviour of many
particles is an obvious next target.
When many particles are confined in \NY{quasi-one-dimensional} channel, they
show collective drift and oscillatory motion \citep{Ikura:2013}.
Similar behaviours is reproduced by the modified model, in which fusion
does not occur.
We will study details of the model in future.
 }

 \NY{
Our \SYYYY{treatment of interaction} is similar to the works in
\citep{Golovin:1995,Lavrenteva:1999}, although there are several differences.
All these models consider the interactions between two spherical
objects that are producing chemical components on their surfaces.
They take into account a boundary condition on the surface and evaluate the interaction by investigating the motion  due to the concentration overlap and hydrodynamics.
In their first attempt \citep{Golovin:1995}, the objects do not undergo
self-propulsive motion.
This corresponds to $\tau_c \rightarrow \infty$ in
(\ref{collision.theory.u}), and thus to ${\bf u}={\bf u}_c + {\bf u}_h$
although there is an additional first-order chemical reaction in their
model and our model includes a damping term of the
 chemicals to describe a buffering effect, which regularizes the
 expansion in our analysis.
The main difference is that our approach uses a diffuse-interface model.
When the drop domains move, this is easier to solve numerically than is the boundary-value problem.
Another advantage is that our method does not rely on axisymmetry, and
therefore, it can be easily extended to the non-axisymmetric case.
In fact, all of the terms in the hydrodynamic interaction are anisotropic, as
demonstrated in the first term of (\ref{two.drop.uh}).
Although the dominant interaction term for concentration overlap is
isotropic, as in (\ref{two.drop.uc}), the higher-order terms are
anisotropic because of the coupling between the relative position and
the deformation.
The disadvantage of our approach is its lack of accuracy; because there is a finite width
at the interface, we are not able to accurately measure the size of the
drop.
In addition, the near-field interaction is so far computed only by using
a boundary-value approach using bispherical coordinates
\citep{Golovin:1995} or by a lubrication analysis.
 Despite these limitations, we believe that our approach provides
 useful  insights to the problem of self-propulsive drops.
 }
 
 We limited ourselves to the cases \NY{in which the} deformation of the drop is
 not large.
 We may relax this assumption by changing the parameters and \NY{we
 expect that this would reveal  intriguing dynamics due to} the coupling between self-propulsion and
 deformation.
 \NY{
 We leave this as a subject for future study.
 }
 This work focuses on head-on \NY{collisions} in detail and suggests the
 importance of concentration-mediated interactions.
 However, there are other types of collision, \NY{such as motion that is
 not parallel to the
 centreline between two drops.
 In these cases, it is possible that the hydrodynamics play a role.
As described by the reduced equations
 (\ref{collision.theory.x})-(\ref{two.drop.uh}), the dominant term in a
 concentration-mediated interaction is isotropic for each drop, while for
 a hydrodynamic interaction, it is not.
 When the centreline is not along the
 direction of the motion, that is, when the incident angle is between 0
 and $\pi/2$, the anisotropic interaction results in the rotation of
 the drops.
 \NYY{
 Our preliminary results suggest that the hydrodynamics
 play a relevant role when the steady velocity is high and the
 deformation of the drop occurs.
 In the current model, fusion occurs at the high steady velocity.
 Nevertheless, by inhibiting fusion, we have
 also obtained the bound state, that is, the state in which two drops
 move together following a collision at a certain incident angle.
 }
 This cannot be reproduced without considering the hydrodynamic interaction.
 The investigation of these motions is \SSY{an} important area for
 future research.
}

\section*{Acknowledgements}
The authors are grateful to Kei-Ichi Ueda and
 Tanniemola Liverpool for
 helpful discussions.
SY acknowledges the support by Grants-in-Aid for Japan Society for
 Promotion of Science (JSPS) Fellows (Grants Nos. 241799 and 263111) and
 the JSPS Core-to-Core Program "Non-equilibrium dynamics of soft matter
 and information".
 The authors acknowledges the support by JSPS KAKENHI Grant Numbers
 JP15K17737 for SY, and JP26800219, JP26103503, and JP16H00793 for NY.

\appendix

\section{Spherical Bessel function}

In this work, we use the spherical Bessel function defined \NY{as}
\begin{align}
j_n(x) 
&=
\sqrt{\frac{\pi}{2x}}
\mathcal{J}_{n+1/2}(x)
\end{align}
where $\mathcal{J}_{n}(x)$ is the $n$-th order Bessel function of
\NY{the} first kind.
The spherical Bessel functions \NY{can also be} expressed in the following way:
\begin{align}
j_n(x) 
&=
(-1)^n x^n 
\left(
\frac{1}{x} \odiff{}{x}
\right)^n
\frac{\sin x}{x}
\label{sphbessel}
\end{align}
The spherical Bessel function satisfies the following relation
\begin{align}
j'_n(x)
&=
\frac{n}{x} j_n(x) - j_{n+1} (x)
\end{align}
where $ j'_n(x) = d j_n(x)/dx$.
For $n=0$, it becomes
\begin{align}
j'_0(x) 
&=
-j_1(x).
\label{besseljp1j0}
\end{align}

\NY{We} consider the following integral\NY{, which contains three spherical
Bessel functions:}
\begin{align}
&
\int_0^{\infty} 
\frac{q^m}{q^2 + \beta^2}
j_l (q R_0)
j_{l'} (q r_{12})
j_{l''} (q R_0)
dq
\nonumber \\
= &
(-1)^{l+l'+l''}
R_0^{l+l''}
r_{12}^{l'}
\left(
\frac{1}{R_0} \pdiff{}{R_0}
\right)^{l}
\left(
\frac{1}{r_{12}} \pdiff{}{r_{12}}
\right)^{l'}
\left(
\frac{1}{R_0} \pdiff{}{R_0}
\right)^{l''}
\int_0^{\infty}
\frac{\sin (q R_0) \sin (q r_{12}) \sin (q R_0)}
{R_0^{2} r_{12}(q^2 + \beta^2)q^{l+l'+l''+3-m}}
dq
\label{integral3sbesselj}
\end{align}
Since $l+l'+l''$ is even and $m$ is either $m=0$ or $m=2$ , the integral does not change under the
transformation $q \rightarrow -q$.
 \NY{We} consider the integral
\begin{align}
I 
&=
\frac{1}{2}
\int_{-\infty}^{\infty} 
\frac{\sin (q R_0) \sin (q r_{12}) \sin (q R_0)}
{(q^2 + \beta^2)q^{l+l'+l''+3-m}}
dq
\nonumber \\
&=
-\frac{1}{16 i}
\int_{-\infty}^{\infty} 
\frac{(e^{i q R_0} - e^{-i q R_0})^2 (e^{i q r_{12}} - e^{-i q r_{12}}) }
{(q^2 + \beta^2)q^{l+l'+l''+3-m}}
dq.
\end{align}
This integral is calculated from residues $q=0,\pm i \beta$.
The main contribution arises from the residue $q=i\beta$ for the
integration path passing $+i\infty$ in the positive direction, and from the residue $q=-i\beta$ for the
integration path passing $-i\infty$ in the negative direction.
For $r_{12} > 2 R_0$,
\begin{align}
I 
&=
- \frac{\pi}{8 }
\left[
\lim_{q \rightarrow i\beta}
\frac{(e^{i q R_0} - e^{-i q R_0})^2 e^{i q r_{12}}  }
{(q + i\beta)q^{l+l'+l''+3-m}}
- (-1)
\lim_{q \rightarrow -i\beta}
\frac{(e^{i q R_0} - e^{-i q R_0})^2 e^{-i q r_{12}} }
{(q - i \beta)q^{l+l'+l''+3-m}}
\right]
\nonumber \\
&=
- \frac{\pi}{8 }
\left[
\frac{4 \sinh^2 (\beta R_0) e^{- \beta r_{12}}  }
{2i\beta (i \beta)^{l+l'+l''+3-m}}
+
\frac{4 \sinh^2 (\beta R_0) e^{- \beta r_{12}} }
{( - 2i \beta) (-i\beta)^{l+l'+l''+3-m}}
\right]
\nonumber \\
&=
- \frac{\pi}{8 }
\frac{ \sinh^2 (\beta R_0) e^{- \beta r_{12}}  }
{i\beta (i \beta)^{l+l'+l''+3-m}}
\label{integral3sbesselj2}
\end{align}

Next, we consider the following general integral
\begin{align}
&
\int_0^{\infty} 
\frac{q^m}{(q^2 + \beta^2)^n}
j_l (q R_0)
j_{l'} (q r_{12})
j_{l''} (q s)
dq
\nonumber \\
= &
(-1)^{l+l'+l''}
R_0^{l}
r_{12}^{l'}
s^{l''}
\left(
\frac{1}{R_0} \pdiff{}{R_0}
\right)^{l}
\left(
\frac{1}{r_{12}} \pdiff{}{r_{12}}
\right)^{l'}
\left(
\frac{1}{s} \pdiff{}{s}
\right)^{l''}
\int_0^{\infty}
\frac{\sin (q R_0) \sin (q r_{12}) \sin (q s)}
{R_0 r_{12} s (q^2 + \beta^2)^n q^{l+l'+l''+3-m}}
 dq
 .
\label{integralRr12sbesselj}
\end{align}
We calculate the following integral
\begin{align}
I_n 
&=
\frac{1}{2}
\int_{-\infty}^{\infty} 
\frac{\sin (q R_0) \sin (q r_{12}) \sin (q s)}
{(q^2 + \beta^2)^n q^{l+l'+l''+3-m}}
dq
\nonumber \\
&=
-\frac{1}{16 i}
\int_{-\infty}^{\infty} 
\frac{(e^{i q R_0} - e^{-i q R_0}) (e^{i q r_{12}} - e^{-i q r_{12}}) (e^{i q s} - e^{-i q s}) }
{(q^2 + \beta^2)^n q^{l+l'+l''+3-m}}
dq.
\end{align}
For $r_{12} > s >  R_0$, we obtain
\begin{align}
I_n
=&
- \frac{\pi}{8}
\left[
\lim_{q \rightarrow i\beta}
\odiff{^{n-1}}{q^{n-1}}
\frac{(e^{i q R_0} - e^{-i q R_0}) (e^{i q s} - e^{-i q s}) e^{i q r_{12}}  }
 {(q + i\beta)q^{l+l'+l''+3-m}}
 \right.
 \nonumber \\
 &
 \left.
- (-1)
\lim_{q \rightarrow -i\beta}
\odiff{^{n-1}}{q^{n-1}}
\frac{(e^{i q R_0} - e^{-i q R_0}) (e^{i q s} - e^{-i q s}) e^{-i q r_{12}} }
{(q - i \beta)q^{l+l'+l''+3-m}}
 \right]
 .
\end{align}

\bibliographystyle{jfm}

\end{document}